\documentclass[twocolumn,american,australian,aps,pra]{revtex4-2}
\usepackage[T1]{fontenc}
\usepackage[utf8]{inputenc}
\usepackage{float}
\usepackage{booktabs}
\usepackage{amsmath}
\usepackage{graphicx}
\usepackage{wasysym}

\makeatletter

\providecommand{\tabularnewline}{\\}

\makeatother

\usepackage{babel}
\begin{document}
\title{\selectlanguage{american}%
Validation tests of Gaussian boson samplers with photon-number resolving
detectors}
\author{\selectlanguage{american}%
Alexander S. Dellios, Margaret D. Reid and Peter D. Drummond}
\email{peterddrummond@protonmail.com}

\affiliation{\selectlanguage{american}%
Centre for Quantum Science and Technology Theory, Swinburne University
of Technology, Melbourne 3122, Australia}
\begin{abstract}
An important challenge with the current generation of noisy, large-scale
quantum computers is the question of validation. Does the hardware
generate correct answers? If not, what are the errors? This issue
is often combined with questions of computational advantage, but it
is a fundamentally distinct issue. In current experiments, complete
validation of the output statistics is generally not possible because
it is exponentially hard to do so. Here, we apply phase-space simulation
methods to partially verify recent experiments on Gaussian boson sampling
(GBS) implementing photon-number resolving (PNR) detectors. The positive-P
phase-space distribution is employed, as it uses probabilistic sampling
to reduce complexity. It is $\sim10^{18}$ times faster than direct
classical simulation for experiments on $288$ modes where quantum
computational advantage is claimed. When combined with binning and
marginalization to improve statistics, multiple validation tests are
efficiently computable, of which some tests can be carried out on
experimental data. We show that the data as a whole shows discrepancies
with theoretical predictions for perfect squeezing. However, a small
modification of the GBS parameters greatly improves agreement for
some tests. We suggest that such validation tests could form the basis
of feedback methods to improve GBS experiments.
\end{abstract}
\maketitle

\section{Introduction}

Theoretical and experimental interest in photonic networks employed
as quantum computers has grown considerably \citep{KnillLaflammeMilburn,AaronsonArkhipov2013LV}.
These devices interfere photons from nonclassical states in a large
scale interferometric network, then measure the resulting output photons
using photodetectors. Of the proposed network architectures, Gaussian
boson sampling (GBS) has proved scalable to the sizes needed to infer
quantum advantage, which recent experiments have now claimed \citep{zhong2020quantum,zhongPhaseProgrammableGaussianBoson2021,madsenQuantumComputationalAdvantage2022,deng2023gaussian}.
In this case, photons are prepared in a squeezed state and the computational
task is sampling from an output distribution with probabilities in
a \#P-hard computational class \citep{Hamilton2017PhysRevLett.119.170501,quesada2018gaussian}.
For photon-number resolving (PNR) detectors this is a Hafnian, while
for threshold detectors it is a Torontonian. Neither function is computable
in polynomial time, so that the random counts cannot be classically
replicated at large scale.

A crucial question with all current generation quantum computers is
how to test for errors in the computed outputs. An efficient and scalable
method for validating GBS with threshold detectors, demonstrated for
up to $16,000$ modes, was proposed in Ref.\citep{drummondSimulatingComplexNetworks2022}
and used to analyze data from a $100$-mode experiment \citep{zhong2020quantum}.
The normally ordered positive-P phase-space representation was utilized
to simulate grouped count probabilities (GCPs). This is a generalized
binned photon counting distribution that includes all measurable marginal
and binned correlations. The exponential sparsity of the detected
count pattern probabilities means the full distribution cannot be
either computed or measured. Therefore, validation tests based on
marginals and binned probability distributions are the most practical
method for validation, as is the case for other random number generators
\citep{Rukhin2010}.

In this paper, we use positive P-representation simulations to test
the validity of Gaussian boson samplers with photon number resolving
(PNR) detectors, extending the tests that can be applied to experimental
data. This is essential, as a number of large scale GBS experiments
have claimed quantum advantage using either PNR \citep{madsenQuantumComputationalAdvantage2022}
or threshold detectors \citep{zhong2020quantum,zhongPhaseProgrammableGaussianBoson2021,deng2023gaussian}.
To perform the tests, we generalize a binning method developed by
Mandel \citep{mandelCoherencePropertiesOptical1965,Mandel1995_book}
for simulating photon statistics of classical light. The original
formulation, which utilized the Glauber-Sudarshan P-representation
\citep{Glauber_1963_P-Rep,Sudarshan_1963_P-Rep} to simulate moments
for classical fields, is extended to the positive P-representation,
which can treat any quantum state. An extension to binning in multiple
dimensions is also provided, as these are more sensitive tests than
one-dimensional distributions.

We show that in the challenging prospect of using quantum computers
to demonstrate quantum computational advantage, the GBS approach has
a very important advantage. Unlike many other proposed quantum computing
architectures, there are multiple quantitative statistical tests that
are dependent on the experimental parameters, which can be used to
validate the output distributions of the data. Experimental data may
fail such validation due to errors, potentially allowing classical
simulations of comparable accuracy. This is an important criterion,
similar to tests of classical random number generators and cryptographic
security \citep{bassham2010sp,knuth2014art}. Implementing a test
suite requires algorithms that compute the correlations or moments
without requiring the full \#P hard distribution.

Like phase-space methods for threshold detectors \citep{drummondSimulatingComplexNetworks2022},
the PNR method is efficient and scalable. Increased dimension of binning
causes an increase in computation time, where in the limit of dimension
equaling mode number the GCP converges to the Hafnian. PNR phase-space
methods allow one to perform simulations of either the ground truth
distribution or modifications of this distribution that include decoherence,
measurement errors and shot-to-shot fluctuations. Such errors are
typically not analyzed, despite evidence suggesting that a thermalized,
non-ideal ground truth distribution provides stronger claims of quantum
advantage than the ideal ground truth distribution \citep{dellios2025validation}.
Unlike threshold detection, PNR detector binning has the advantage
of not requiring the computation of multidimensional inverse discrete
Fourier transforms, which can increase computation time for increased
dimension.

The positive-P phase-space distribution is exact, non-singular and
positive for any quantum state. The moments of the distribution are
identical to normally ordered operator moments including intensity
correlations \citep{Glauber1963_CoherentStates}, which are reproduced
to any order. This property, coupled with probabilistic random sampling,
makes the positive P-representation ideal for validating the statistics
of nonclassical state photon counting experiments such as GBS. It
does not have the computational restraints imposed by other validation
methods \citep{villalonga2021efficient,ohSpoofingCrossEntropyMeasure2023,bulmerBoundaryQuantumAdvantage2022a},
since it is many orders of magnitude faster than direct classical
simulation, with negligible rounding errors. While there are sampling
errors, for current experimental parameters they are comparable to
sampling errors in the measurement statistics. Both can be reduced
in a controllable way.

Other validation methods involve comparing experimental count patterns
to those obtained from classical algorithms that replicate the photo-detection
measurements either exactly \citep{bulmerBoundaryQuantumAdvantage2022a}
or approximately \citep{villalonga2021efficient,ohSpoofingCrossEntropyMeasure2023}.
Exact algorithms generate count patterns by directly sampling the
distributions, but are limited to small experiments only. Approximate
algorithms typically neglect high-order correlations, and only use
the low-order correlations. Despite the usefulness of these algorithms
in defining classical simulation boundaries of quantum advantage and
determining the correlations present in a GBS experiment, they are
either computationally demanding or lack high-order correlations,
which is necessary for claims of quantum advantage. This limits the
application of these approaches when validating GBS experiments and
investigating physical processes arising within networks.

In order to determine the validity of using the positive P-representation
to simulate photon counting experiments with PNR detectors, we compare
simulated GCP moments to the exactly known photon counting distributions
for special cases. Here, multi-mode pure squeezed states are transformed
by a Haar random unitary matrix which is multiplied by a uniform amplitude
loss coefficient. For losses comparable to current experiments, we
see excellent agreement. In the less realistic lossless case, the
resulting non-Gaussian number distribution oscillates strongly between
even and odd total counts. While this is far from current experiments,
the standard positive P-representation requires a much larger number
of phase-space samples to converge in such cases. However, this limit
can also be efficiently simulated if necessary using a symmetry-projected
\citep{drummond2016coherent} phase-space method, which will be treated
elsewhere.

Finally, using photon count data from the recent Borealis experiment
implementing PNR detectors \citep{madsenQuantumComputationalAdvantage2022},
we apply the simulated GCPs to validate the moments of measured photon
counting distributions by comparing theory and experiment. Differences
are quantified using chi-square and Z-statistic tests. As with earlier
work \citep{drummondSimulatingComplexNetworks2022,dellios2025validation},
we show that the experimental GCPs are far from the ground truth distribution
for pure squeezed state inputs. Instead, they are described much better
by a different target distribution which includes decoherence and
measurement error corrections of the transmission matrix. Although
we focus specifically on implementations to GBS, we emphasize that
one can use the positive P-representation to simulate GCPs of any
linear photon counting experiment using classical or nonclassical
state photons.

This paper is structured as follows, Sections \ref{sec:GCP_theory}
and \ref{sec:Normally-ordered-phase-space-met} introduce the necessary
background and theoretical methods for simulating GCPs of any photon
counting experiment using PNR detectors in phase-space. Section \ref{sec:GBS_sample_requirments}
then applies this method to the exactly known multi-mode squeezed
state distributions, while Section \ref{sec:GBS_GCP_exp_comp} simulates
the theoretical distributions corresponding to experimental data from
the Borealis GBS claiming quantum advantage. Finally, \ref{sec:Conclusion}
summarizes the main findings of this paper.

\section{Grouped count probabilities\label{sec:GCP_theory}}

We first consider a standard GBS experiment where $N$ single-mode
squeezed vacuum states are sent into a linear photonic network defined
by an $M\times M$ transmission matrix $\boldsymbol{T}$. If each
input mode is independent, the full input density operator is defined
as
\begin{equation}
\hat{\rho}^{(\text{in})}=\prod_{j}^{N}\left|r_{j}\right\rangle \left\langle r_{j}\right|,\label{eq:input_density_operator}
\end{equation}
where $\left|r_{j}\right\rangle =\hat{S}(r_{j})\left|0\right\rangle $
denotes the squeezed vacuum state of the $j$-th mode with squeezing
parameter $r_{j}$ and $\hat{S}(r_{j})$ is the squeezing operator
\citep{Drummond2004_book}.

Since squeezed states are Gaussian, they are characterized by their
quadrature variances
\begin{align}
\left\langle :\left(\Delta\hat{x}_{j}\right)^{2}:\right\rangle  & =2\left(n_{j}+m_{j}\right)=e^{2r_{j}}-1\nonumber \\
\left\langle :\left(\Delta\hat{y}_{j}\right)^{2}:\right\rangle  & =2\left(n_{j}-m_{j}\right)=e^{-2r_{j}}-1,\label{eq:pure_squeezed_state_quad_variance}
\end{align}
where the quadrature operators $\hat{x}_{j}=\hat{a}_{j}^{(\text{in})}+\hat{a}_{j}^{\dagger(\text{in})}$
and $\hat{y}_{j}=\left(\hat{a}_{j}^{(\text{in})}-\hat{a}_{j}^{\dagger(\text{in})}\right)/i$
satisfy the commutation relations $\left[\hat{x}_{j},\hat{y}_{k}\right]=2i\delta_{jk}$,
and the mean photon number and coherence per mode is defined as $n_{j}=\left\langle \hat{a}_{j}^{\dagger(\text{in})}\hat{a}_{j}^{(\text{in})}\right\rangle =\sinh^{2}(r_{j})$
and $m_{j}=\left\langle \hat{a}_{j}^{(\text{in})}\hat{a}_{j}^{(\text{in})}\right\rangle =\cosh(r_{j})\sinh(r_{j})$,
respectively. Throughout this paper, we use $:\cdots:$ to denote
normal ordering.

While most theoretical analyses of GBS consider pure squeezed state
inputs, this is impractical, since decoherence effects such as mode
mismatches \citep{zhong2020quantum}, partial distinguishability \citep{shiEffectPartialDistinguishability2022},
and many more \citep{drummond2020initial} will always be present
during experimental applications in quantum optics. Therefore, a realistic
model of input states are Gaussian thermalized squeezed states for
which a simple parameterization was introduced in \citep{drummondSimulatingComplexNetworks2022}.
Here, a thermalization component $\epsilon$ alters the intensity
of the input photons as measured by the coherence, which is reduced
by a factor of $1-\epsilon$ such that $\tilde{m}_{j}=(1-\epsilon)m_{j}$.
The quadrature variances are altered as \citep{drummondSimulatingComplexNetworks2022}
\begin{align}
\sigma_{x_{j}}^{2}=\left\langle :\left(\Delta\hat{x}_{j}\right)^{2}:\right\rangle  & =2\left(n_{j}+\tilde{m}_{j}\right)\nonumber \\
\sigma_{y_{j}}^{2}=\left\langle :\left(\Delta\hat{y}_{j}\right)^{2}:\right\rangle  & =2\left(n_{j}-\tilde{m}_{j}\right),\label{eq:thermal_squeezed_state_quad_variance}
\end{align}
where for $\epsilon=0$, the pure squeezed state definition Eq.(\ref{eq:pure_squeezed_state_quad_variance})
returns, while $\epsilon=1$ produces fully decoherent thermal states
with $\sigma_{x_{j}}^{2}=\sigma_{y_{j}}^{2}=2n_{j}$. More complex
thermalized models are also possible, for example with mode dependent
thermalization, $\epsilon_{j}$.

In the ideal GBS scenario, the transmission matrix is a Haar random
unitary. However experimentally realistic transmission matrices contain
losses, making them non-unitary. Throughout this paper, $\boldsymbol{T}$
is used to define a lossy transmission matrix, while $\boldsymbol{U}$
defines a Haar random unitary. When losses are present, the creation
and annihilation operators for the input modes $\hat{a}_{j}^{\dagger(\text{in})},\hat{a}_{j}^{(\text{in})}$
are converted to output modes following 
\begin{equation}
\hat{a}_{k}^{\dagger(\text{out})}=\sum_{j=1}^{N}T_{kj}\hat{a}_{j}^{\dagger(\text{in})}+\sum_{j=1}^{M}L_{kj}\hat{b}_{j}^{\dagger(\text{in})},\label{eq:lossy_input_output_relationship}
\end{equation}
where $\boldsymbol{L}$ is a random noise matrix representing photon
loss to $j$-the environmental mode $\hat{b}_{j}^{\dagger(\text{in})}$.
The resulting output photon number $\hat{n}'_{k}=\hat{a}_{k}^{\dagger(\text{out})}\hat{a}_{k}^{(\text{out})}$
is then measured by a series of photodetectors.

\subsection{Photon counting projectors}

Detectors with photon number resolution implement the normally ordered
projection operator \citep{walls2008quantum,Sperling2012True}
\begin{equation}
\hat{p}_{j}(c_{j})=\frac{1}{c_{j}!}:\left(\hat{n}'_{j}\right)^{c_{j}}e^{-\hat{n}'_{j}}:,\label{eq:PNR_detector_projector}
\end{equation}
where $c_{j}=0,1,2,3,\dots$ is the number of measured photon counts
at the $j$-th detector. Although theoretically there is no upper
bound on $c_{j}$, practically PNR detectors saturate for some maximum
number of counts $c_{j}^{(\text{max})}$. This value is variable,
typically depending on the physical implementation of the detector.

In any multi-mode photon counting experiment, a set of photon count
patterns $\boldsymbol{c}=[c_{1},\dots,c_{M}]$ are observed by measuring
the output state $\hat{\rho}^{(\text{out})}$. A straightforward extension
of Eq.(\ref{eq:PNR_detector_projector}) allows one to define the
projection operator for a specific count pattern as
\begin{equation}
\hat{P}(\boldsymbol{c})=\bigotimes_{j=1}^{M}\hat{p}_{j}(c_{j}).
\end{equation}
For GBS, each pattern is a sample from the full output probability
distribution and computing its expectation value corresponds to solving
the matrix Hafnian \citep{Hamilton2017PhysRevLett.119.170501,kruse2019detailed}
\begin{equation}
\left\langle \hat{P}(\boldsymbol{c})\right\rangle =\frac{1}{\sqrt{\text{det}(\boldsymbol{Q})}}\frac{\left|\text{Haf}\left(\boldsymbol{B}_{S}\right)\right|^{2}}{\prod_{j=1}c_{j}!},
\end{equation}
which is a \#P-hard computational task due to its relationship to
the matrix permanent. Here, $\text{Haf}(\boldsymbol{B}_{S})$ is the
Hafnian of the square sub-matrix of $\boldsymbol{B}=\boldsymbol{U}\left(\bigoplus_{j=1}^{M}\tanh(r_{j})\right)\boldsymbol{U}^{T}$
for pure state inputs with zero displacement, where the sub-matrix
is formed from the output channels with recorded photon counts. The
symmetric matrix $\boldsymbol{B}$ is related to the kernel matrix
$\boldsymbol{A}=\boldsymbol{B}\bigoplus\boldsymbol{B}^{*}$ which
is formed from the $2M\times2M$ covariance matrix $\boldsymbol{Q}$
as \citep{quesada2020exact}
\begin{equation}
\boldsymbol{A}=\left(\begin{array}{cc}
0 & \boldsymbol{I}_{M}\\
\boldsymbol{I}_{M} & 0
\end{array}\right)\left(\boldsymbol{I}_{M}-\boldsymbol{Q}^{-1}\right),
\end{equation}
where $\boldsymbol{I}_{M}$ denotes the $M\times M$ identity matrix.
When the covariance matrix elements are non-negative, the kernel matrix
and hence the symmetric matrix $\boldsymbol{B}$ are also non-negative
\citep{quesada2020exact} and the Hafnian can be estimated in polynomial
time \citep{barvinokPolynomialTimeAlgorithms1999a,rudelsonHafniansPerfectMatchings2016}.

In the current generation of GBS networks, photon loss is the dominant
noise source. Although there is evidence that sampling from a lossy
GBS distribution is also a \#P-hard computational task \citep{deshpandeQuantumComputationalAdvantage2022a},
output photons from experiments implementing high loss networks become
classical \citep{qi2020regimes}, allowing the output distribution
to become efficiently sampled. There is currently no clear boundary
between the computationally hard and efficiently sampled regimes,
but a classical algorithm that exploits large experimental photon
loss has recently been developed \citep{oh2024classical} and shown
to produce samples that are closer to the ideal ground truth than
recent experiments claiming quantum advantage. The effects of thermalization
on the computational complexity of GBS is currently unknown, despite
thermalization causing photons from nonclassical states to become
classical at a faster rate than pure losses.

\subsection{Grouped count probabilities for PNR detectors}

Determining photon counting distributions by binning the photons arriving
at photodetectors has a long history in quantum optics. Although binning
methods can vary, e.g. binning continuous measurements within a time
interval such as inverse bandwidth time \citep{huangPhotoncountingStatisticsMultimode1989,zhuPhotocountDistributionsContinuouswave1990},
such distributions are both experimentally accessible and theoretically
useful. For pure squeezed states these can provide evidence of nonclassical
behavior, as observed in the even-odd count oscillations \citep{Mehmet2010Observation}.
For GBS there is also a statistical purpose, as the set of all possible
observable patterns $\mathcal{S}_{P}$ grows exponentially with mode
number, causing the output probabilities to become exponentially sparse.

In this paper, we are interested in the total output photon number
operator obtained from a subset of $S_{j}$ efficient detectors:
\begin{equation}
\hat{n}'_{S_{j}}=\sum_{i\in S_{j}}\hat{n}'_{i}.\label{eq:total_output_photon_number}
\end{equation}
Our observable of interest is then the photon number distribution
corresponding to counting the projectors onto eigenstates of $\hat{n}'_{S_{j}}$.

Following from the nomenclature introduced in \citep{drummondSimulatingComplexNetworks2022},
we define the $d$-dimensional grouped photon counting distribution
as the probability of observing $\boldsymbol{m}=[m_{1},\dots,m_{d}]$
grouped counts in $j=1,\dots,d$ distinct subsets of output modes
$\boldsymbol{S}=\text{\ensuremath{\left(S_{1},\dots,S_{d}\right)}}$,
such that 
\begin{equation}
\mathcal{G}_{\boldsymbol{S}}^{(n)}(\boldsymbol{m})=\left\langle \prod_{j=1}^{d}\left[\sum_{\sum c_{i}=m_{j}}\hat{P}_{S_{j}}(\boldsymbol{c})\right]\right\rangle ,\label{eq:GCP_PNR}
\end{equation}
where $n=\sum_{j=1}^{d}M_{j}\leq M$ is the total correlation order
and each grouped count is obtained from the summation $m_{j}=\sum_{i\in S_{j}}c_{i}$.
The general form of the pattern projector for any number of output
modes within each subset is 
\begin{equation}
\hat{P}_{S_{j}}\left(\boldsymbol{c}\right)=\bigotimes_{i\in S_{j}}\frac{1}{c_{i}!}:(\hat{n}'_{i})^{c_{i}}e^{-\hat{n}'_{i}}:.
\end{equation}

The correlation order and dimension determines the type of observable
one is simulating. If $n=M$, all correlation orders are simulated
in phase-space. Meanwhile $n<M$ denotes a marginal probability, since
the remaining $M-n$ output modes are ignored. If $n=M$ and $d=1$,
the GCP becomes the univariate distribution $\mathcal{G}_{S}^{(M)}(m)$
where all modes are contained within a single subset $S=\{1,\dots,M\}$.
This is called the total count distribution as it's the probability
of observing $m$ counts in any pattern and is identical to the total
photon-number distribution of Ref. \citep{madsenQuantumComputationalAdvantage2022}.

When $d>1$, multi-dimensional GCPs are simulated. These are joint
probability distributions over $\boldsymbol{m}$ counts. For example,
one of the $d=2$ GCPs is the bivariate photon count distribution
$\mathcal{G}_{S_{1},S_{2}}^{(M)}(m_{1},m_{2})$, which is the probability
of observing $m_{1}$ counts in $S_{1}=\{1,\dots,M_{1}\}$, and $m_{2}$
counts in $S_{2}=\{M_{1}+1,\dots,M\}$. The largest possible dimension
is $d=M$ where the vector of subsets becomes $\boldsymbol{S}=(S_{1},S_{2},\dots,S_{M})=(\{1\},\{2\},\dots,\{M\})$.
In this case, the GCP to converges to the Hafnian as the joint probabilities
have become pattern probabilities.

Provided one has the computational resources (see Section \ref{sec:Numerical-scaling}),
comparisons of multi-dimensional GCPs provide a number of benefits
over their $d=1$ counterparts. The first is the number of count bins
scales as $\mathcal{B}=(Md^{-1}+1)^{d}$, allowing more detailed comparisons
with theory and experiment. The second is by randomly permuting the
measured count patterns, an exponentially large number of comparison
tests can be performed on experimental data, where each permutation
allows one to compare different binned correlation moments \citep{dellios2025validation}.

So long as $n=M$ all correlation orders are simulated in phase-space.
However, since GCP probabilities converge to the Hafnian as $d\rightarrow M$,
a greater range of distinct correlations are statistically significant
in this limit. If these correlations are nonclassical, multi-dimensional
GCPs provide a more thorough test of quantum advantage, because they
allow one to directly compare more sums of high-order correlations
with their theoretical values \citep{dellios2025validation}, and
these tests are more sensitive to the exact squeezing and transfer
matrix parameter values.

\section{Normally-ordered phase-space methods\label{sec:Normally-ordered-phase-space-met}}

Here we summarize phase-space methods, and explain the motivation
for using them. The purpose is to transform an exponentially complex
Hilbert space into a positive distribution that can be randomly sampled.
Normal-ordering methods are the closest to experimental statistics
using photon-counting. Representations that are not normally ordered,
like the Wigner \citep{Wigner_1932} or Q-function \citep{Husimi1940}
approach, have a rapidly growing sampling error with system size.

\subsection{Glauber-Sudarshan representation}

In its original formulation, Mandel \citep{mandelCoherencePropertiesOptical1965,Mandel1995_book}
used the normally ordered Glauber-Sudarshan P-representation \citep{Glauber_1963_P-Rep,Sudarshan_1963_P-Rep}
to evaluate the expectation value in Eq.(\ref{eq:GCP_PNR}). The P-representation
expands the density operator as a diagonal sum of coherent states
\citep{Glauber_1963_P-Rep}
\begin{equation}
\hat{\rho}=\int P(\boldsymbol{\alpha})\left|\boldsymbol{\alpha}\right\rangle \left\langle \boldsymbol{\alpha}\right|\text{d}^{2M}\boldsymbol{\alpha},
\end{equation}
hence its commonly called the diagonal P-representation, where $\left|\boldsymbol{\alpha}\right\rangle =\bigotimes_{j=1}^{M}\left|\alpha_{j}\right\rangle $
is a multi-mode coherent state vector with eigenvalues $\boldsymbol{\alpha}=[\alpha_{1},\dots,\alpha_{M}]^{T}$
and $P(\boldsymbol{\alpha})$ is the P-representation probability
distribution. The diagonal P-representation can be used to evaluate
expectation values of the product of normally ordered operators due
to the relationship
\begin{align}
\left\langle \prod_{j=1}^{M}:\hat{a}_{j}^{\dagger}\hat{a}_{j}:\right\rangle  & =\lim_{E_{S}\rightarrow\infty}\left\langle \prod_{j=1}^{M}\left|\alpha_{j}\right|^{2}\right\rangle _{DP}\nonumber \\
 & =\int P(\boldsymbol{\alpha})\left(\prod_{j=1}^{M}\left|\alpha_{j}\right|^{2}\right)\text{d}^{2M}\boldsymbol{\alpha},
\end{align}
where $\left\langle \dots\right\rangle $ is a quantum expectation
value, $E_{S}$ is the total number of samples, and $\left\langle \dots\right\rangle _{DP}$
is the diagonal-P phase-space ensemble mean.

Following the standard replacement of $\hat{a}_{j}\rightarrow\alpha_{j}$
and $\hat{a}_{j}^{\dagger}\rightarrow\alpha_{j}^{*}$, where $n'_{j}=\left|\alpha_{j}\right|^{2}$
and $\hat{n}'_{S_{j}}\rightarrow n'_{S_{j}}=\sum_{i\in S_{j}}\left|\alpha_{i}\right|^{2}$,
Mandel \citep{mandelCoherencePropertiesOptical1965,Mandel1995_book}
simulates Eq.(\ref{eq:GCP_PNR}) for $d=1$ and $n=M$ as 
\begin{equation}
\mathcal{G}_{S}^{(M)}(m)=\int P(\boldsymbol{\alpha})\left[\frac{1}{m!}\left(n'_{S}\right)^{m}e^{-n'_{S}}\right]\text{d}^{2M}\boldsymbol{\alpha}.
\end{equation}
Since the density operator of nonclassical states contains off-diagonal
elements due to quantum superpositions, if the diagonal P-representation
is used for simulations of nonclassical states, $P(\boldsymbol{\alpha})$
becomes singular and negative. From standard mathematical requirements
that it should be real, normalizable, non-negative and non-singular,
it is no longer a probability distribution. This method is therefore
only useful for classical states such as coherent, thermal and squashed
states \citep{MartinezCifuentes2023classicalmodelsmay,Reid1986},
since these have a positive diagonal P-distribution.

\subsection{Positive-P representation}

To simulate the photon number distributions of nonclassical states
one must instead use the normally ordered positive P-representation
\citep{Drummond_generalizedP1980}, which is part of a family of generalized
P-representations that produce non-singular and exact phase-space
distributions for any quantum state. The positive P-representation
expands the density operator as a $4M$-dimensional volume integral
\citep{Drummond_generalizedP1980}
\begin{equation}
\hat{\rho}=\iint P(\boldsymbol{\alpha},\boldsymbol{\beta})\frac{\left|\boldsymbol{\alpha}\right\rangle \left\langle \boldsymbol{\beta}\right|}{\left\langle \boldsymbol{\beta}|\boldsymbol{\alpha}\right\rangle }\text{d}^{2M}\boldsymbol{\alpha}\text{d}^{2M}\boldsymbol{\beta},\label{eq:+P_representation_general}
\end{equation}
where the additional dimensions allow off-diagonal density matrix
elements to be included in the coherent state basis $\left|\boldsymbol{\beta}\right\rangle =\bigotimes_{j=1}^{M}\left|\beta_{j}\right\rangle $
with eigenvalues $\boldsymbol{\beta}=[\beta_{1},\dots,\beta_{M}]^{T}$.
The coherent amplitudes $\boldsymbol{\alpha}$, $\boldsymbol{\beta}$
are independent and can vary along the entire complex plane. We can
also take the real part of Eq.(\ref{eq:+P_representation_general}),
since the density matrix is hermitian. The benefit of using the positive
P-representation for photon counting experiments is that it includes
the diagonal P-representation when the positive-P distribution satisfies
$P(\boldsymbol{\alpha},\boldsymbol{\beta})=P(\boldsymbol{\alpha})\delta(\text{\ensuremath{\boldsymbol{\alpha}}-\ensuremath{\boldsymbol{\beta}}})$.
This allows us to define classical phase-space as having $\boldsymbol{\beta}=\boldsymbol{\alpha}$,
while a nonclassical phase space requires non-vanishing probability
of $\boldsymbol{\beta}\neq\boldsymbol{\alpha}$.

Expectation values of products of normally ordered operators for nonclassical
states can now be directly calculated from the moments of the positive-P
distribution 
\begin{align}
\left\langle \prod_{j=1}^{M}:\hat{a}_{j}^{\dagger}\hat{a}_{j}:\right\rangle  & =\lim_{E_{S}\rightarrow\infty}\left\langle \prod_{j=1}^{M}\alpha_{j}\beta_{j}^{*}\right\rangle _{P}\nonumber \\
 & =\iint P(\boldsymbol{\alpha},\boldsymbol{\beta})\left(\prod_{j=1}^{M}\alpha_{j}\beta_{j}^{*}\right)\text{d}^{2M}\boldsymbol{\alpha}\text{d}^{2M}\boldsymbol{\beta},\label{eq:+P_moment_exptation_value_relationship}
\end{align}
where we define $\left\langle \dots\right\rangle _{P}$ as the positive-P
ensemble mean and have used the c-number replacements $\hat{a}_{j}\rightarrow\alpha_{j}$
and $\hat{a}_{j}^{\dagger}\rightarrow\beta_{j}^{*}$.

To simulate Eq.(\ref{eq:GCP_PNR}) for nonclassical states requires
replacing the total output photon number Eq.(\ref{eq:total_output_photon_number})
by the phase-space observable 
\begin{equation}
n'_{S_{j}}=\sum_{i\in S_{j}}n'_{i}=\sum_{i\in S_{j}}\alpha'_{i}\left(\beta'_{i}\right)^{*},
\end{equation}
such that the general multidimensional probability distribution is
\begin{align}
\mathcal{G}_{\boldsymbol{S}}^{(n)}(\boldsymbol{m}) & =\left\langle \prod_{j=1}^{d}\left[\sum_{\sum c_{i}=m_{j}}P_{S_{j}}(\boldsymbol{c})\right]\right\rangle _{P},\nonumber \\
 & =\text{Re}\int P(\boldsymbol{\alpha},\boldsymbol{\beta})\left[\prod_{j=1}^{d}\left[\frac{1}{m_{j}!}\left(n'_{S_{j}}\right)^{m_{j}}e^{-n'_{S_{j}}}\right]\right]\text{d}\mu,\label{eq:GCP_PNR_+P_general}
\end{align}
where $\text{d}\mu\equiv\text{d}\boldsymbol{\alpha}\text{d}\boldsymbol{\beta}$
and the phase-space pattern projector is
\begin{equation}
P_{S_{j}}(\boldsymbol{c})=\prod_{i\in S_{j}}\frac{1}{c_{i}!}(n'_{i})^{c_{i}}e^{-n'_{i}}.
\end{equation}
In Eq.(\ref{eq:GCP_PNR_+P_general}), the required summation over
output photon number is incorporated into the pattern projector, allowing
one to make the substitution $m_{j}=\sum_{i\in S_{j}}c_{i}$, while
we take the real part as required for pure squeezed state applications
presented in the next subsection.

Equation \ref{eq:GCP_PNR_+P_general} is a more general definition
of GCPs than Mandel's original formulation, since one can now simulate
the marginal and multidimensional probabilities of any classical or
nonclassical photon counting experiment. To simulate GCPs for input
photons in a number state, it is more efficient to use another generalized
P-representation: the complex P-representation. In this case $P(\boldsymbol{\alpha},\boldsymbol{\beta})$
includes a complex weight due to expanding the density operator as
a contour integral \citep{Drummond_generalizedP1980}. Although the
positive P-representation exists for Fock states, it has larger sampling
errors than the complex version \citep{opanchuk2018simulating}, meaning
fewer stochastic samples are needed to perform accurate simulations.

Unlike methods for threshold detectors \citep{drummondSimulatingComplexNetworks2022},
Eq.(\ref{eq:GCP_PNR_+P_general}) doesn't require multidimensional
inverse discrete Fourier transforms, which can increase the computation
time. Since threshold detector outputs are always binary, the largest
number of observable photon counts in a pattern corresponds to a detection
event in every output mode, such that $\sum_{i}^{M}c_{i}\leq M$ where
the largest observable grouped count is $m^{(M)}$.

Since PNR detectors can resolve multiple photons arriving at each
detector, an experiment or classical algorithm replicating photo-detection
measurements can now produce count patterns where $\sum_{i}^{M}c_{i}>M$,
in which case $m^{(M)}$ is no longer the largest observable grouped
count. Instead of scaling with mode number, numerical simulations
scale with photon count bins, where we define the maximum observable
grouped count as $m^{(\text{max})}$. For experiments with small mean
input photon numbers or high losses, one may have $m^{(\text{max})}<m^{(M)}$,
while low losses and large photon numbers will produce $m^{(\text{max})}>m^{(M)}$.

Typically, we choose $m^{(\text{max})}$ to occur when probabilities
are of the order $\mathcal{G}_{S}^{(n)}(m^{(\text{max})})\leq10^{-7}$,
providing a standard cutoff for very small probabilities as was implemented
in previous methods \citep{drummondSimulatingComplexNetworks2022,dellios2025validation}.
A minimum observable grouped count $m^{(\text{min})}$ can also be
introduced to form an analogous cut-off $\mathcal{G}_{S}^{(n)}(m^{(\text{min})})\leq10^{-7}$.
This allows our algorithm to only simulate grouped counts within the
relevant range $m_{j}^{(\text{min})}\leq m_{j}\leq m_{j}^{(\text{max})}$,
reducing memory use and computation times.

\subsection{Squeezed states in phase-space}

Initial coherent amplitudes are generated by randomly sampling the
phase-space distribution of a specific state. For most states, analytical
forms of this distribution exist, allowing one to derive an analytical
expression for the initial samples of each state. Here we simply recall
previously derived results for squeezed states \citep{drummondSimulatingComplexNetworks2022},
so that the theory presented is complete, since the form of the positive-P
distribution doesn't depend on the type of measurement, that is, PNR
or not.

For applications to GBS, we first consider the ideal case with pure
squeezed state inputs. In order to define the input density operator
Eq.(\ref{eq:input_density_operator}) in the positive P-representation
we first expand each state as an integral over real coherent states
\citep{adam1994complete}
\begin{equation}
\left|r_{j}\right\rangle =\sqrt{C_{j}}\int\exp\left(\frac{-\alpha_{j}^{2}}{\gamma_{j}}\right)\left|\alpha_{j}\right\rangle \text{d}\alpha_{j},
\end{equation}
with normalization constant $C_{j}=\sqrt{\gamma_{j}+1}/(\pi\gamma_{j})$
and $\gamma_{j}=\exp(2r_{j})-1$. After some straightforward calculations,
the input density operator becomes
\begin{equation}
\hat{\rho}^{(\text{in})}=\text{Re}\iint P(\boldsymbol{\alpha},\boldsymbol{\beta})\frac{\left|\boldsymbol{\alpha}\right\rangle \left\langle \boldsymbol{\beta}\right|}{\left\langle \boldsymbol{\beta}|\boldsymbol{\alpha}\right\rangle }\text{d}\boldsymbol{\alpha}\text{d}\boldsymbol{\beta},
\end{equation}
where the positive-P distribution for pure squeezed states is \citep{drummondSimulatingComplexNetworks2022}
\begin{equation}
P(\boldsymbol{\alpha},\boldsymbol{\beta})=\prod_{j}C_{j}e^{-\left(\alpha_{j}^{2}+\beta_{j}^{2}\right)\left(\gamma^{-1}+1/2\right)+\alpha_{j}\beta_{j}}.
\end{equation}

Thermalized squeezed states, defined by the $\epsilon$ component
introduced \foreignlanguage{australian}{in Section \ref{sec:GCP_theory}},
are readily incorporated into the positive-P distribution. A general
random sampling formalism is used to generate initial stochastic samples
for any Gaussian input \citep{drummondSimulatingComplexNetworks2022}
\begin{align}
\alpha_{j} & =\frac{1}{2}\left(\sigma_{x_{j}}w_{j}+i\sigma_{y_{j}}w_{j+M}\right)\nonumber \\
\beta_{j}^{*} & =\frac{1}{2}\left(\sigma_{x_{j}}w_{j}-i\sigma_{y_{j}}w_{j+M}\right),\label{eq:+P_initial_samples}
\end{align}
where $\left\langle w_{j}w_{k}\right\rangle _{P}=\delta_{jk}$ are
real Gaussian noises and the variances $\sigma_{x_{j}}$, $\sigma_{y_{j}}$
are defined in Eq.(\ref{eq:thermal_squeezed_state_quad_variance}).
For classical inputs, $\sigma_{y_{j}}$ is real and $\alpha_{j}=\beta_{j}$.
For quantum inputs, $\sigma_{y_{j}}$ is imaginary and $\alpha_{j}\neq\beta_{j}$.
Following from Eq.(\ref{eq:lossy_input_output_relationship}), the
output stochastic amplitudes used to perform simulations are then
obtained via the transformations $\boldsymbol{\alpha}'=\boldsymbol{T}\boldsymbol{\alpha}$
and $\boldsymbol{\beta}'=\boldsymbol{T}\boldsymbol{\beta}$ where
the amplitudes $\boldsymbol{\alpha}'$, $\boldsymbol{\beta}'$ correspond
to measurements of the density matrix $\hat{\rho}^{\text{(out)}}$.

\section{Exact models of GBS grouped count distributions\label{sec:GBS_sample_requirments}}

In this section, we validate the sampled positive-P method by comparing
the results to exactly known distributions of GBS networks whose size,
mean squeezing parameters and average loss rates correspond to those
found in the PNR experiments of Madsen et al \citep{madsenQuantumComputationalAdvantage2022}.
This demonstrates that for the relevant regime, the phase-space sampling
errors are negligible. A full comparison with experiment that uses
the inhomogeneous parameters describing the squeezing vector and transmission
matrix is carried out in Section \ref{sec:GBS_GCP_exp_comp}.

The positive-P algorithm used throughout this paper is part of the
phase-space network simulation program xqsim, which is available from
the Github repository \citep{GitHubPeterddrummondXqsim}.

\subsection{Exactly known distributions}

The photon statistics of pure squeezed states have been studied extensively
in quantum optics \citep{huangPhotoncountingStatisticsMultimode1989,zhuPhotocountDistributionsContinuouswave1990,Hamilton2017PhysRevLett.119.170501,deshpandeQuantumComputationalAdvantage2022a}.
Assuming the input modes have equal squeezing parameters $r=r_{1}=\dots=r_{N}$
and hence, equal photon numbers $n=n_{1}=\dots=n_{N}$, the one-dimensional
GCP, or total count probability, can be computed exactly. For the
case of pure squeezed states transformed by a matrix with uniform
amplitude loss coefficient $t$, applied as $t\boldsymbol{U}$, the
exact total count distribution takes the form \citep{deshpandeQuantumComputationalAdvantage2022a}
\begin{align}
\mathcal{G}_{S}^{(M)}(2m) & =t^{4m}\mathcal{C}p^{M/2}(1-p)^{m}f_{1/2}\nonumber \\
\mathcal{G}_{S}^{(M)}(2m-1) & =2m(1-t^{2})t^{4m-2}\mathcal{C}p^{M/2}(1-p)^{m}f_{3/2},\label{eq:GCP_exact_uniform_loss_squeezed_state}
\end{align}
which is defined for $m>0$, where $S=\{1,\dots,M\}$, $p=1/(1+n)$
is the success probability, $\mathcal{C}=\binom{\frac{M}{2}+m-1}{m}$
and $f_{c}={}_{2}F_{1}\left(a,b;c;z\right)$ is the Gauss hypergeometric
function with $a=m+\frac{1}{2}$, $b=\frac{M}{2}+m$, $z=(1-t^{2})^{2}(1-p)$.

For a lossless network with $t=1$, $_{2}F_{1}\left(a,b;c;0\right)=1$
and Eq.(\ref{eq:GCP_exact_uniform_loss_squeezed_state}) converges
to the well known distribution \citep{huangPhotoncountingStatisticsMultimode1989,zhuPhotocountDistributionsContinuouswave1990,Hamilton2017PhysRevLett.119.170501}
\begin{align}
\mathcal{G}_{S}^{(M)}(2m) & =\mathcal{C}p^{M/2}(1-p)^{m}\nonumber \\
\mathcal{G}_{S}^{(M)}(2m-1) & =0,\label{eq:GCP_exact_squeezed_state}
\end{align}
which contains distinct oscillations between even and odd counts.
In the limit $M\rightarrow\infty$, it reduces to a Poissonian for
the pair counts \citep{zhuPhotocountDistributionsContinuouswave1990}
\begin{align}
\mathcal{G}_{S}^{(M)}(2m) & =\frac{1}{m!}e^{-Mn/2}\left(\frac{Mn}{2}\right)^{m}\nonumber \\
\mathcal{G}_{S}^{(M)}(2m-1) & =0,
\end{align}
while the full distribution is not Poissonian.

Such photon statistics are characteristic of photon bunching which,
unlike photon anti-bunching, is not an exclusively nonclassical phenomena
\citep{walls2008quantum}. However as squeezed photons are always
generated in highly correlated pairs, bunching is indicative of the
non-classicality of the measured photons. The presence of even-odd
oscillations \citep{Mehmet2010Observation}, even for networks with
decoherence and loss, provides an explicit test of non-classicality
although their absence does not exclude non-classicality.

When even-odd count oscillations are present in lossless and ultra-low
loss photonic networks, the positive-P moments suffer from slow convergence
and large sampling errors. Since the even-odd count oscillations vanish
as soon as one has a high chance of losing a photon, the lossless
and ultra-low loss regimes are not relevant to any current or near
future experimental network, which are both far from lossless and
whose photon number distribution is approximately Gaussian for large
system sizes with loss rates as small as $t=0.95$. Therefore, a detailed
theoretical analysis of the positive P-representation in the lossless
and ultra-low loss regimes will not be performed here, as projected
methods \citep{drummond2016coherent} may converge faster in such
regimes.

\subsection{Comparisons with small and large sized networks}

To test whether Eq.(\ref{eq:GCP_PNR_+P_general}) can be utilized
to simulate nonclassical photon counting experiments using PNR detectors
with perfect squeezers, photodetectors and photonic networks with
uniform photon loss, we first simulate small sized GBS networks $M=N=16$
and $M=N=72$ with uniform squeezing $\boldsymbol{r}=[0.89,\dots,0.89]$
and amplitude loss rates $t=0.6$ and $t=0.56$, respectively.

\begin{figure}
\begin{centering}
\includegraphics[width=0.5\textwidth]{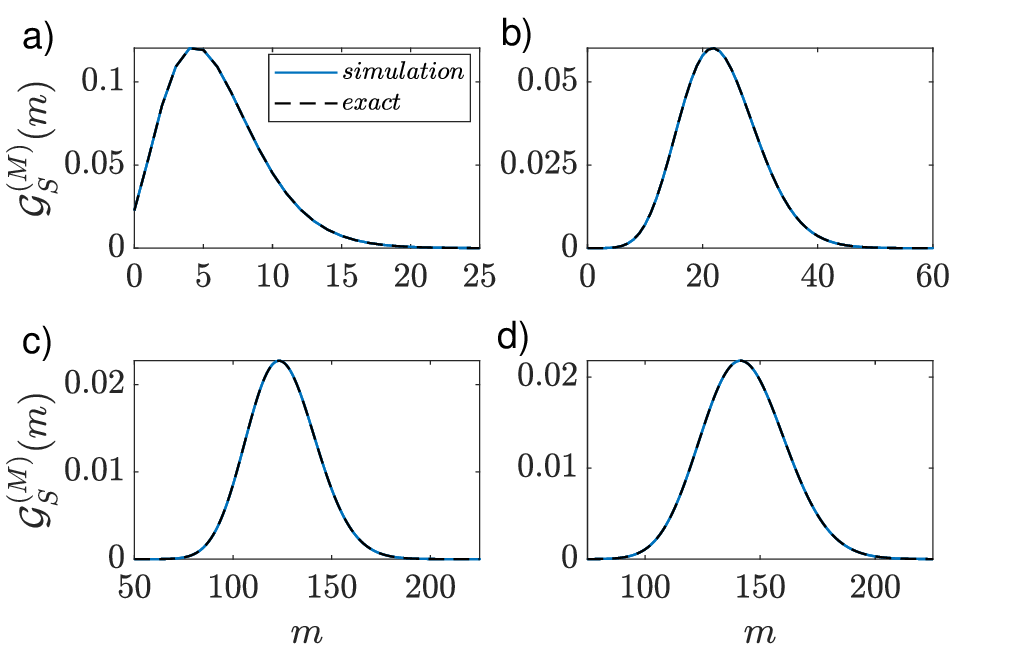}
\par\end{centering}
\caption{\foreignlanguage{australian}{Validation of the positive-P GCP distributions (solid blue lines)
by comparison to exact multi-mode squeezed state photon counting distributions
(dashed black lines) in an exactly soluble example. We use uniform
amplitude loss rates $t$ for different network sizes, squeezing parameters
and loss rates corresponding to the various experimental networks
implemented in Ref.\citep{madsenQuantumComputationalAdvantage2022}.
a) A GBS set-up with $M=N=16$ is considered where pure squeezed states
with uniform squeezing parameter $\boldsymbol{r}=[0.89,\dots,0.89]$
are transformed by a Haar random unitary matrix multiplied by a uniform
amplitude loss coefficient $t\boldsymbol{U}$ with $t=0.6$. Network
sizes are now increased to b) $M=N=72$, with $\boldsymbol{r}=[0.89,\dots,0.89]$
and $t=0.56$, c) $M=N=216$ with $\boldsymbol{r}=[1.1,\dots,1.1]$
and $t=0.57$, and d) $M=N=288$ with $\boldsymbol{r}=[1,\dots,1]$
and $t=0.6$. Positive-P distributions are obtained by averaging over
the ensemble of samples $E_{S}=2.4\times10^{6}$. Upper and lower
lines correspond to $\pm1\sigma_{T,j}$, where $\sigma_{T,j}\propto1/\sqrt{E_{S}}$
is the theoretical sampling errors for the $j$-th photon count bin,
which are so small that they are not visible in the output graphics.
\label{fig:Exact_+P_PNR_comp}}}
\end{figure}

Comparisons between positive-P moments and the exact distribution
Eq.(\ref{eq:GCP_exact_uniform_loss_squeezed_state}) are presented
in Fig.(\ref{fig:Exact_+P_PNR_comp}a) and Fig.(\ref{fig:Exact_+P_PNR_comp}b)
for these networks where simulations are performed for an ensemble
size of $E_{S}=2.4\times10^{6}$. In the limit of large $E_{S}$ the
positive-P moments converge to moments of a distribution (see Eq.(\ref{eq:+P_moment_exptation_value_relationship})).
The exact size of $E_{S}$ required for a given accuracy varies. An
ensemble size of the order $E_{S}\ge10^{6}$ is generally needed for
sampling and difference errors of less than $10^{-3}$, typical of
current experimental accuracy.

For both network sizes and loss rates, the positive-P moments converge
to the required exact distribution with errors $<10^{-3}$. Increasing
the network sizes to $M=N=216$ and $M=N=288$, positive-P moments
simulated for loss rates $t=0.57$ and $t=0.6$ are presented in Fig.(\ref{fig:Exact_+P_PNR_comp}c)
and Fig.(\ref{fig:Exact_+P_PNR_comp}d). In both cases, moments converge
to the exact distribution with errors $<10^{-3}$ such that they are
not visible in the graphics. Therefore, in the loss regimes of the
current experiments, the positive P-representation provides an accurate
method to validate experimental networks. For larger sized networks
within the ultra-low loss regime this is also true, as the even-odd
count oscillations vanish for much smaller amplitude loss rates.

\subsection{Numerical scaling\label{sec:Numerical-scaling}}

For validation tests to be useful, they must be accurate, efficient
and scalable. The results above have validated the accuracy of the
positive P-representation by comparing the simulated distribution
to an exactly known model. Next, we analyze the efficiency and scalability
of simulating GCPs using the positive P-representation for varying
mode number $M$ and dimension $d$.

\selectlanguage{australian}%
When performing phase-space simulations, one can subdivide the total
number of samples into two sub-ensembles \foreignlanguage{american}{$E_{S}=N_{S}N_{R}$.
This has both a statistical benefit, as described in the Appendix,
and a computational one, as it allows both vector and multi-core parallel
computation, drastically reducing computation time.}

\selectlanguage{american}%
The first sub-ensemble, $N_{S}$, is the number of samples generated
from a phase-space distribution, in this case Eq.(\ref{eq:+P_initial_samples}),
corresponding to a specific stochastic trajectory. The number of trajectories
is denoted by second sub-ensemble $N_{R}$.

Generally, the size of each sub-ensemble affects the resulting computation
time. As $N_{S}\rightarrow\infty$, large memory bandwidth to store
vectors and matrices to $64$-bit (double) precision is required.
Meanwhile, using $N_{R}\rightarrow\infty$ is more efficient if the
hardware has more CPU cores. In practice, so long as $E_{S}$ remains
unchanged, one can vary the size of each sub-ensemble to suit hardware
requirements without affecting the overall accuracy of the simulation,
except that the accuracy of sampling error estimates may be reduced
at either extreme.

\begin{figure}
\begin{centering}
\includegraphics[width=0.5\textwidth]{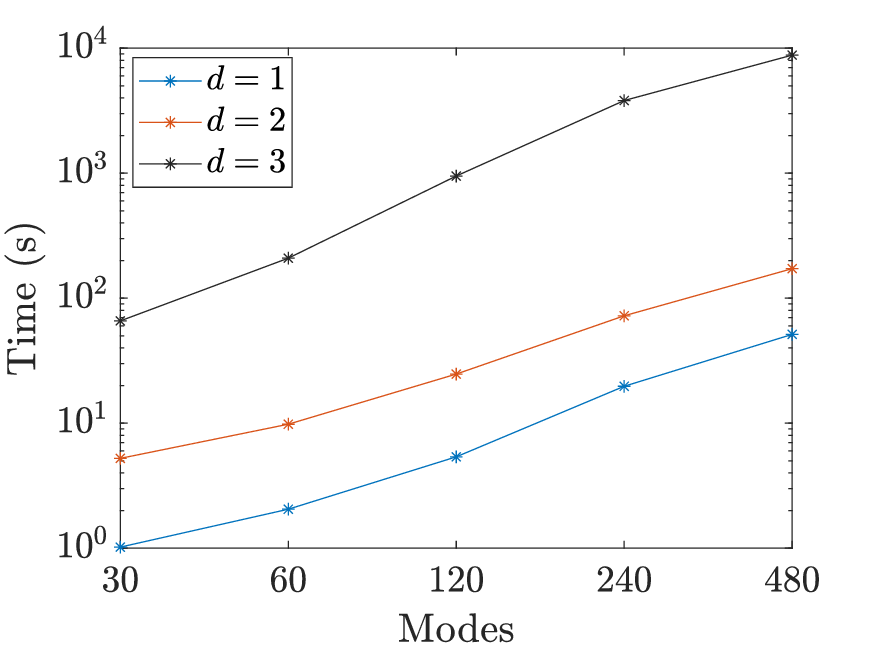}
\par\end{centering}
\caption{\foreignlanguage{australian}{Run-times in seconds versus mode number of positive-P simulated GCPs
of dimension $d=1$ (blue), $d=2$ (orange), and $d=3$ (black). GBS
networks with uniform squeezing parameter\foreignlanguage{american}{
$\boldsymbol{r}=[1,\dots,1]$ and loss rate $t=0.6$ (i.e. the mean
of the parameters used in Fig.(\ref{fig:Exact_+P_PNR_comp})) are
simulated} for \foreignlanguage{american}{mode numbers $M=30,60,\dots,480$}
where the network matrix is a Haar random unitary. The algorithm is
optimised by only simulating counts within $m_{j}^{(\text{min})}\protect\leq m_{j}\protect\leq m_{j}^{(\text{max})}$.
The set of maximum counts for $d=1,2,3$ are $\{62,88,134,218,369\}$,
$\{37,50,80,115,190\}$, and $\{25,33,49,77,126\}$ respectively,
while the set of minimum counts are $\{0,5,12,45,134\}$, $\{0,0,0,15,60\}$,
and $\{0,0,2,11,41\}$. All positive-P simulations are performed using
$E_{S}=2.4\times10^{6}$.\foreignlanguage{american}{ Sub-ensemble
sizes were varied as $N_{S}=500,50,5$ and $N_{R}=4.8\times10^{3},4.8\times10^{4},4.8\times10^{5}$
for $d=1,2,3$ respectively}, in order to optimize run-times on the
utilized hardware. \label{fig:+P_benchmarking}}}
\end{figure}

To investigate the efficiency and scalability of our method, GCPs
of dimension $d=1,2,3$ are simulated for GBS networks $M=N=30,60,\dots,480$
with $\boldsymbol{r}=[1,\dots,1]$ and $t=0.6$. Positive-P numerical
run-times for these networks are presented in Fig.(\ref{fig:+P_benchmarking})
and were obtained using a typical current (2025) desktop computer
with a 12-core 3.7 GHz AMD CPU and 32GB of RAM. All simulations used
$E_{S}=2.4\times10^{6}$ samples, although the sub-ensembles were
varied to optimize run-times such that $N_{S}=500,50,5$ and $N_{R}=4.8\times10^{3},4.8\times10^{4},4.8\times10^{5}$
for $d=1,2,3$ respectively. For all networks and dimensions tested,
sampling errors were $<10^{-3}$.

\selectlanguage{australian}%
From Fig.(\ref{fig:+P_benchmarking}) it is clear that positive-P
run-times scale nearly linearly with mode number for all dimensions.
As the dimension increases, run-times increase by approximately an
order of magnitude. For example, the largest simulated network run-times
(in seconds) are $51.34$, $172.3$, and $8800$ for $d=1,2,\:\text{and}\:3$,
respectively. This is unexpected, since the floating-point operations
needed for the transfer matrix multiplication scale as $M^{2}$, but
it indicates that memory access is more important for this range of
network sizes.

\selectlanguage{american}%

\selectlanguage{australian}%
The sharp jump in time from $d=2$ to $d=3$ is due to increased memory
requirements. The available memory needed to store a single complex
multi-dimensional array to double-precision scales as $16N_{S}(M/d)^{d}$
bytes. Hence when $M=480$, each $d=2$ array is $\approx0.921N_{S}$
MB, while for $d=3$ it is $\approx65.5N_{S}$ MB. Naturally, additional
memory is necessary to actually perform the computations.

Although more sensitive tests of quantum advantage are possible by
comparing $d>1$ GCPs of theory and experiment, there are computational
costs. Such costs are not restricted to phase-space representations.
Hardware limitations are present in all quantum computing validation
schemes. One must therefore balance test sensitivity with available
hardware.
\selectlanguage{american}%

\section{Numerical simulations of experimental networks\label{sec:GBS_GCP_exp_comp}}

In this section, we use the sampled positive-P method to validate
experimental data of the GBS device Borealis \citep{madsenQuantumComputationalAdvantage2022}.
The experimental data corresponds to the publicly available dataset
of Madsen et al \citep{madsenQuantumComputationalAdvantage2022}.
Other works \citep{goldberg2023measuring,stanev2025validation} have
validated independently generated data from the Borealis network.
These have focused on calculating the quadrature coherence scale \citep{goldberg2023measuring}
and orbits \citep{stanev2025validation} of Borealis.

In this work we focus on quantitatively validating GCPs of different
dimensionality, which has not been done previously. Detailed $\chi^{2}$
and Z-statistic tests are applied to compare theory and experiment
for both pure and thermalized squeezed state inputs. Due to claims
of quantum advantage, most of the analysis in this section is on the
count patterns from the $216$-mode high squeezing (HS) parameter
and $288$-mode experimental data sets. Summary tables of $\chi^{2}$
and Z-statistic test results for the Borealis data sets can be found
in Appendix A.

\subsection{Statistical tests}

A number of statistical tests have been proposed to validate experimental
GBS networks. These include total variation distance \citep{villalonga2021efficient,madsenQuantumComputationalAdvantage2022},
cross-entropy measures \citep{villalonga2021efficient,ohSpoofingCrossEntropyMeasure2023,oh2024classical},
two and higher-order correlation functions \citep{phillips2019benchmarking,zhongPhaseProgrammableGaussianBoson2021},
and Bayesian hypothesis testing \citep{zhongPhaseProgrammableGaussianBoson2021,MartinezCifuentes2023classicalmodelsmay}.
Although these tests provide useful insights into the behavior of
specific correlations, they are limited to either small size systems
or patterns with small numbers of detected photons, as they require
computing the probabilities of specific count patterns, which is a
\#P-hard task.

If these patterns are closer to the ground truth distribution than
classical algorithms replicating photo-detection measurements, then
its assumed patterns from larger sized systems claiming quantum advantage
will also pass these tests. The ground truth distribution is the target
distribution GBS aims to sample from when losses are included in the
transmission matrix. Generally its assumed to be close to the ideal
lossless distribution, although this may not necessarily be the case.

In this work, the primary statistical test used to validate experimental
data are chi-square tests \citep{pearson1900x}, which in terms of
GCPs are defined as \citep{drummondSimulatingComplexNetworks2022}
\begin{equation}
\chi^{2}=\sum_{i=1}^{k}\frac{\left(\mathcal{G}_{E,i}-\bar{\mathcal{G}}_{S,i}\right)^{2}}{\sigma_{i}^{2}}.
\end{equation}
Here we use the shorthand notation $\mathcal{\bar{G}}_{i}=\left\langle \mathcal{G}_{\boldsymbol{S}}^{(n)}(m_{j_{i}})\right\rangle _{P}=E_{S}^{-1}\sum_{e=1}^{E_{S}}\left(\mathcal{G}_{\boldsymbol{S}}^{(n)}(m_{j_{i}})\right)_{(e)}$
for the stochastic trajectory $e\in E_{S}$ to denote the positive-P
GCP ensemble means of the $i=1,2,\dots,k$ photon count bin for the
$j=1,\dots,d$ grouped count, which converges to the true theoretical
GCP in the limit $\mathcal{G}_{S,i}=\lim_{E_{S}\rightarrow\infty}\bar{\mathcal{G}}_{S,i}$
for any general input state $S$. Simulated ensemble means are then
compared to the experimental GCPs $\mathcal{G}_{E,i}=N_{E}^{-1}\sum_{i}c_{i}$
formed from $N_{E}\subset\mathcal{S}_{P}$ photon count patterns.

The $\chi^{2}$ tests are used for two reasons: the first is that
this is a standard test applied to data from random number generators
\citep{knuth2014art}, where the summation is performed over $k$
count bins satisfying $\sum_{i\in S_{j}}c_{i}>10$. The second is
that one can include dependence on both theoretical and experimental
sampling errors as $\sigma_{i}^{2}=\sigma_{E,i}^{2}+\sigma_{T,i}^{2}$,
where $\sigma_{E,i}\propto1/\sqrt{N_{E}}$ is the estimated experimental
sampling error. This provides a method of accurately determining whether
observed differences are caused by large sampling error, as is the
case for unbinned sparse data.

Output $\chi^{2}$ values follow a distribution that is positively
skewed for small $k$. Performing a Wilson-Hilferty transformation
\citep{wilsonDistributionChiSquare1931}, which transforms the result
as $\chi^{2}\rightarrow\left(\chi^{2}/k\right)^{1/3}$, the resulting
distribution converges to a Gaussian distribution $\mathcal{N}\left(\mu_{\mathcal{N}},\sigma_{\mathcal{N}}^{2}\right)$
with mean $\mu_{\mathcal{N}}=(9k-2)/9k$ and variance $\sigma_{\mathcal{N}}^{2}=2/(9k)$
for $k\geq10$ due to the central limit theorem \citep{wilsonDistributionChiSquare1931,johnsonContinuousUnivariateDistributions1970}.
This allows one to convert $\chi^{2}$ results into measurable probability
distances using approximate Z-statistic, or Z-score, tests \citep{freundStatisticalMethods2003},
where
\begin{equation}
Z=\frac{\left(\chi^{2}/k\right)^{1/3}-\mu_{\mathcal{N}}}{\sigma_{\mathcal{N}}}.
\end{equation}
This measures how many standard deviations away a test statistic is
from its expected normally distributed mean. The Z-statistic therefore
determines the probability of obtaining the observed result. Agreement
within sampling errors produces test results of $\chi^{2}/k\approx1$
and $\left|Z\right|\leq1$ when accounting for stochastic fluctuations
in the output statistics. For Z-statistic tests we use a regime of
validity with the limit $Z\leq6$ to define accurate comparisons.
Above this limit test statistics differ by more than $6\sigma_{\mathcal{N}}$
from their predicted means, corresponding to very small observable
probabilities.

The results presented in subsequent subsections use the notation introduced
previously \citep{dellios2025validation}. Comparisons between theory
and experiment for simulations with pure squeezed states input into
the experimental $M\times M$ $\boldsymbol{T}$-matrix are denoted
by the subscript $EI$, which we call the ideal ground truth distribution.
Our ideal ground truth corresponds to the ground truth distributions
used in Ref. \citep{madsenQuantumComputationalAdvantage2022}. This
differs from the non-ideal, or modified, ground truth distribution
that includes more realistic experimental effects such as thermalization,
and measurement corrections in the theoretical simulations, which
are denoted by the subscript $ET$.

The validation tests implemented in this paper focus on comparing
experimental data with two theoretical ground truth distributions.
If experimental data fail the statistical tests implemented here,
it does not necessarily mean there is no possible quantum advantage
present in these experiments, but it does mean that the quantum computer
has generated the wrong answer relative to the target ground truth
distribution.

For example, a result of $Z_{EI}\gg1$ or $Z_{ET}\gg1$ for GCPs of
dimension $d>1$ indicates high-order correlations present in experimental
data are far from their expected values for both tested ground truths.
Since quantum advantage requires nonclassical correlations at high-orders,
this result may indicate the tested experiment is dominated by low-order
correlations, possibly increasing the likelihood of classical sampling
algorithms producing samples that are closer to either distribution.
The larger the binning dimension, the more sensitive the tests are
to these correlations, making them harder to pass. However, at too
large a dimension the experimental sampling error increases, reducing
the power of the test to discriminate between different distributions.

To determine whether or not there is any quantum advantage requires
comparisons with classical sampling algorithms that have a comparable
or smaller error. Previous work \citep{dellios2025validation} has
performed some of these validation tests for a single sampling algorithm
compared to threshold detector experiments. We save a comparable analysis
for the PNR experiments considered in this paper for future work.

\subsection{$d=1$ grouped count comparisons}

\begin{figure}
\begin{centering}
\includegraphics[width=0.5\textwidth]{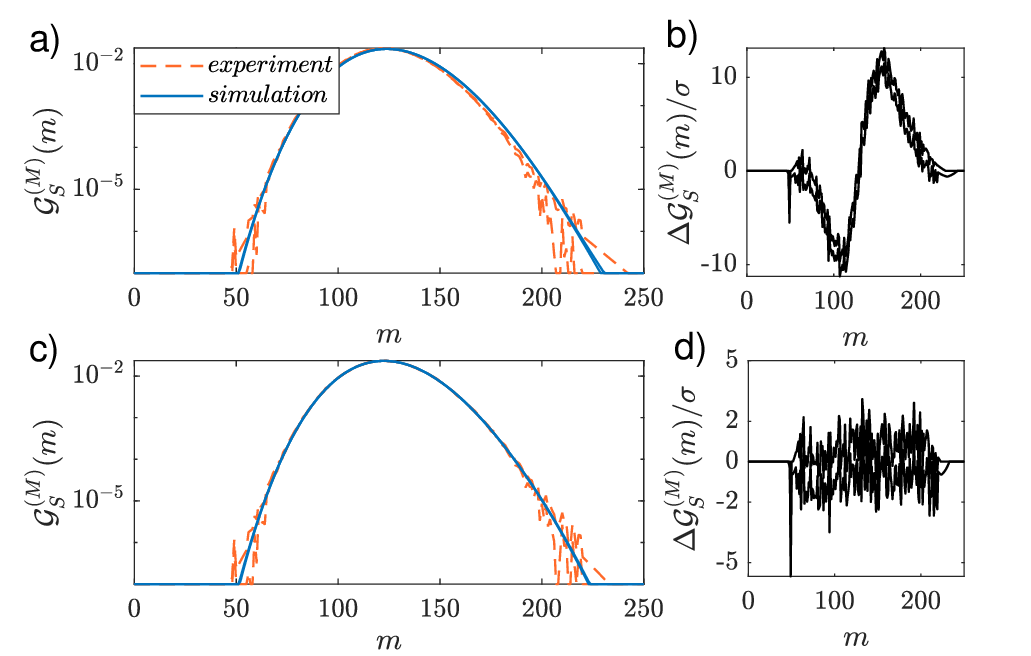}
\par\end{centering}
\caption{\foreignlanguage{australian}{Comparisons of one-dimensional GCPs $\mathcal{G}_{\{1,\dots,M\}}^{(M)}(m)$
versus grouped counts $m$ for experimental data obtained from the
$216$-mode HS Borealis experiment. a) Data is compared to positive-P
simulations with pure squeezed state inputs, where simulations are
performed for an ensemble size of $E_{S}=1.2\times10^{6}$, and the
corresponding normalised difference between distributions is plotted
in b). In c) experimental patterns are now compared to the simulated
GCP using thermalised squeezed state inputs with $\epsilon=0.0510\pm0.0005$.
A transmission matrix measurement correction is also applied as $t\boldsymbol{T}$,
where $t=0.9941\pm0.0005$, to improve the fit. d) The normalised
difference between thermalized and experimental GCPs, which display
noticeable improvement when compared to pure squeezed state simulations.
\label{fig:216_HS_total_counts_combined}}}
\end{figure}

\begin{figure}
\begin{centering}
\includegraphics[width=0.5\textwidth]{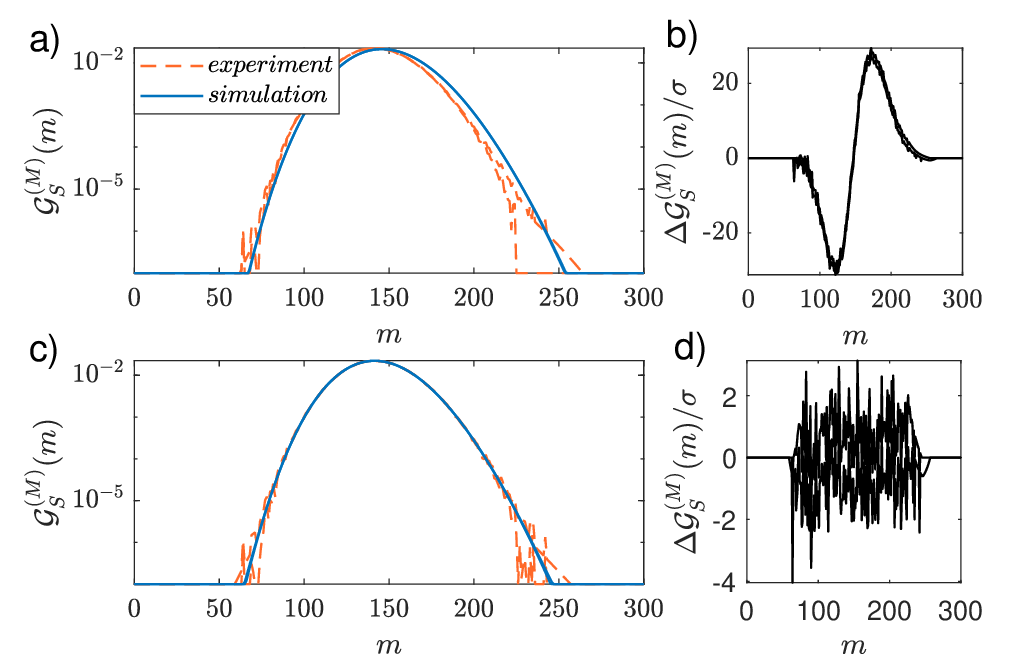}
\par\end{centering}
\caption{\foreignlanguage{australian}{Description follows Fig.(\ref{fig:216_HS_total_counts_combined})
except experimental count patterns are from the $288$-mode Borealis
network. Input thermalised squeezed state simulations are performed
using a decoherence component of $\epsilon=0.0547\pm0.0005$ and matrix
measurement error $t=0.9848\pm0.0005$. \label{fig:288_total_counts_combined}}}
\end{figure}

First we compare the theoretical and experimental one-dimensional
GCPs of the ideal ground truth distribution. Results of phase-space
simulations are presented in Fig.(\ref{fig:216_HS_total_counts_combined}a)
for the $216$-mode HS data set and Fig.(\ref{fig:288_total_counts_combined}a)
for the $288$-mode data set. In both cases, the simulated GCP ideal
ground truth distributions appear to visually resemble the experimental
data.

Visual comparisons of these distributions were also presented in \citep{madsenQuantumComputationalAdvantage2022}.
When comparing our theoretical total count distributions to those
obtained in \citep{madsenQuantumComputationalAdvantage2022}, we found
that for the $288$-mode data set, the two theoretical distributions
differed slightly. After further checks by the original authors \citep{PrivateCommunicationPhoton},
the theoretical total count distribution in Fig.(\ref{fig:288_total_counts_combined}a)
was determined to be the correct ideal ground truth distribution.

Despite the apparent visual agreement between theory and experiment,
a quantitative $\chi^{2}$ test shows significant differences between
the two distributions. The $216$-mode HS experimental data gives
a result of $\chi_{EI}^{2}/k\approx54$ with $Z_{EI}\approx70$ .
Count patterns from the $288$-mode experiment have $\chi_{EI}^{2}/k\approx411$
with $Z_{EI}\approx167$. Although both experimental GCPs are far
from their pure squeezed state distributions, the $288$-mode distribution
is $98\sigma_{\mathcal{N}}$ further from its expected normally distributed
mean. These numerical differences are also reflected graphically by
plotting the normalized difference between distributions $\Delta\mathcal{G}_{S}^{(M)}(m)/\sigma_{i}=\left(\mathcal{G}_{E,i}-\bar{\mathcal{G}}_{I,i}\right)/\sigma_{i}$,
which is plotted in Fig.(\ref{fig:216_HS_total_counts_combined}b)
for the $216$-mode HS data set and Fig.(\ref{fig:288_total_counts_combined}b)
for the $288$-mode data set.

Once a thermal component $\epsilon$ is added to the input states
and a transmission matrix measurement correction $t$ is applied as
$t\boldsymbol{T}$, the positive-P simulated GCPs agree with the experimental
distributions within sampling error. The $216$-mode HS data set requires
$\epsilon=0.0510\pm0.0005$ and $t=0.9941\pm0.0005$ to output $\chi_{ET}^{2}/k\approx0.88\pm0.02$
and $\left|Z_{ET}\right|\approx1\pm0.2$, while the best-fit theoretical
GCP for the $288$-mode data uses a similar thermal fraction of $\epsilon=0.0547\pm0.0005$
and measurement correction $t=0.9848\pm0.0005$ to produce $\chi_{ET}^{2}/k\approx0.87\pm0.02$
and $Z_{ET}\approx\left|1.1\right|\pm0.2$. This improved agreement
between theory and experiment is reflected in the normalized difference
plots as seen in Fig.(\ref{fig:216_HS_total_counts_combined}d) and
Fig.(\ref{fig:288_total_counts_combined}d) which have noticeably
decreased from their pure squeezed state comparisons.

Similar results are obtained for the $72$-mode and $216$-mode low
squeezing (LS) experiments. In these cases, comparisons with the ideal
ground truth, which output $Z_{EI}=72$ and $Z_{EI}=66$ respectively,
converge to a within sampling error agreement with a thermalized theory
such that $\left|Z_{ET}\right|\approx0.2\pm0.2$ for the $72$-mode
experiment and $\left|Z_{ET}\right|\approx0.89\pm0.1$ for the $216$-mode
LS experiment.

\subsection{$d=2$ grouped count comparisons}

\begin{figure}
\begin{centering}
\includegraphics[width=0.5\textwidth]{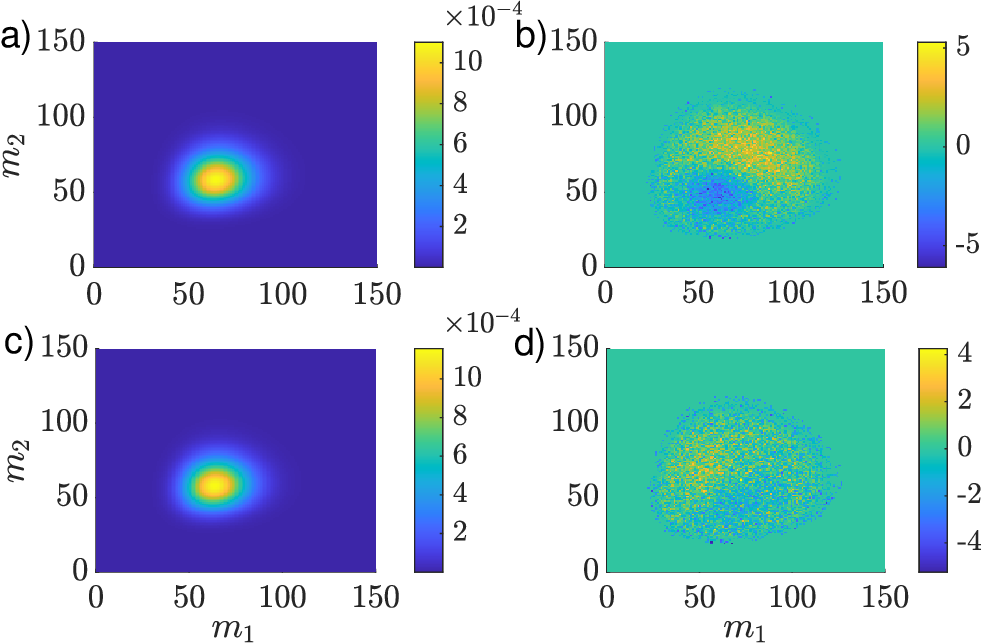}
\par\end{centering}
\caption{\foreignlanguage{australian}{Surface plots of two-dimensional GCP comparisons of theory and experiment
for the $216$-mode HS experimental count patterns. Figures in the
left column (a and c) plot of the full bivariate distribution $\mathcal{G}_{\{1,\dots,108\},\{109,\dots,216\}}^{(M)}(m_{1},m_{2})$,
while the figures in the right column (b and d) plot the normalised
difference $\Delta\mathcal{G}_{\{1,\dots,108\},\{109,\dots,216\}}^{(M)}(m_{1},m_{2})/\sigma$.
In a) and b) the theoretical ideal ground truth is simulated, and
in c) and d) the theoretical distribution is the thermalized ground
truth. Thermalization component $\epsilon$ and measurement correction
$t$, as well as the positive-P sample size are the same as in Fig.(\ref{fig:216_HS_total_counts_combined}).
\label{fig:216_HS_2D_combined}}}
\end{figure}

\selectlanguage{australian}%
To determine whether this excellent agreement between\foreignlanguage{american}{
theoretical and experimental GCPs once decoherence and measurement
corrections are accounted for continues, we increase the binning dimension
to $d=2$. In this case, the $216$-mode HS experiment, whose ideal
and thermalized $d=2$ distributions are shown in Fig.(\ref{fig:216_HS_2D_combined}),
outputs statistical test results of $Z_{EI}=58$ and $Z_{ET}=9\pm1$.}

\selectlanguage{american}%
Although the $Z_{ET}$ results are $\approx8\sigma_{\mathcal{N}}$
larger than their $d=1$ counterparts, the ideal ground truth comparisons
are $\approx12\sigma_{\mathcal{N}}$ closer to the two-dimensional
distribution than the total count comparisons, presumably because
of the smaller number of counts per bin. For the $72$-mode, $216$-mode
LS and $288$-mode experiments, this isn't necessarily the case (see
Appendix A), however all are closer to the thermalized ground truth
than was observed in previous work \citep{dellios2025validation}
with threshold detector experiments.

In these threshold detector experiments, $N_{E}\approx4\rightarrow5\times10^{7}$
count patterns were generated and it was found that once the experimental
data was binned in $d=4$ dimensions, there were too few counts per
bin. This causes the bins to become more sparse, leading to a dramatic
increase in experimental sampling errors that dominate statistical
testing, outputting artificially small results. Therefore for these
threshold detector experiments, reliable validation tests could only
be performed for $d<4$ \citep{dellios2025validation}.

Almost all data sets from Borealis generated only $N_{E}\approx1\times10^{6}$
patterns. In this case, sampling errors dominate statistical testing
at $d=2$, reducing the reliability of these tests such that only
the $d=1$ comparisons are reliable.

The exception to this is the $16$-mode experiment whose $N_{E}\approx8.4\times10^{7}$
samples were used benchmark the system architecture through total
variation distance tests. The network is small enough that count pattern
probabilities can be computed up to a total of six detected photons.
As the computational cost of computing probabilities grows with the
total photon number, the full photon distribution was not computed
in the analysis of Ref.\citep{madsenQuantumComputationalAdvantage2022}.

Interestingly, the one-dimensional GCP for this data set is further
from its ideal ground truth distribution than most experiments performed
using the Borealis quantum network, where $\chi_{EI}^{2}/k\approx2.5\times10^{3}$
and $Z_{EI}=163$. Unlike data sets claiming quantum advantage, including
decoherence and $\boldsymbol{T}$-matrix corrections is no longer
enough to obtain agreement within sampling error between theoretical
and experimental GCPs, instead producing statistical test results
of $\chi_{ET}^{2}/k\approx14\pm2$ and $Z_{ET}\approx18\pm1$, although
the difference here is mostly from the increased sample numbers.

\begin{figure}
\begin{centering}
\includegraphics[width=0.5\textwidth]{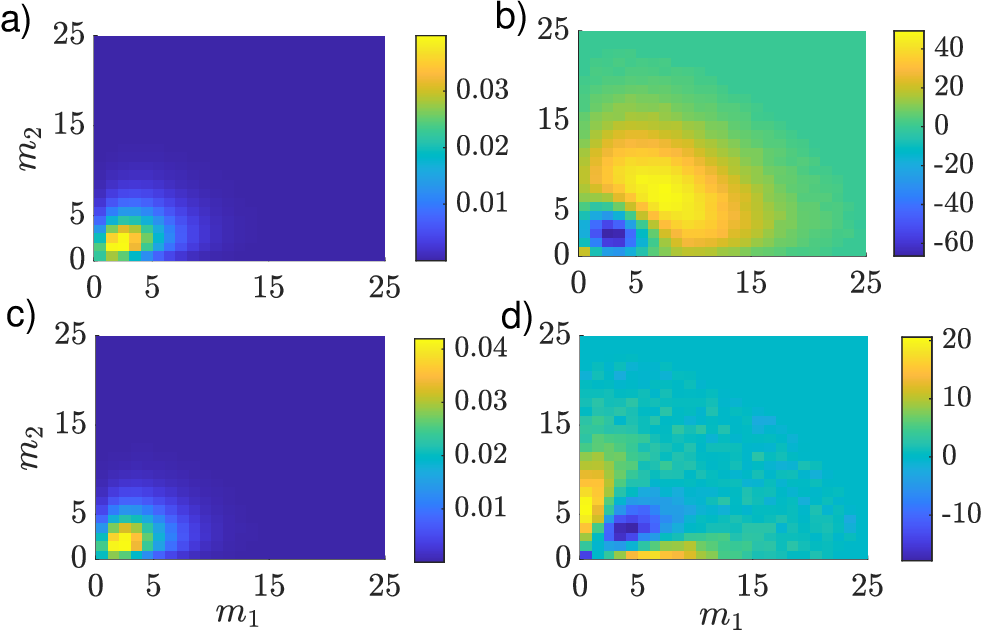}
\par\end{centering}
\caption{\foreignlanguage{australian}{Description follows Fig.(\ref{fig:216_HS_2D_combined}) however in
this case experimental count patterns are from the $16$-mode Borealis
network, such that the $d=2$ GCP is $\mathcal{G}_{\{1,\dots,8\},\{9,\dots,16\}}^{(M)}(m_{1},m_{2})$.
For this network, the thermalization component is $\epsilon=0.0648\pm0.0005$
while the matrix measurement error is $t=0.9841\pm0.0005$. All simulations
used $E_{S}=1.2\times10^{6}$ samples.\label{fig:16M_2D_combined}}}
\end{figure}

An increase in the GCP dimension to $d=2$ shows the $16$-mode data
is even further from both the ideal and non-ideal ground truth distributions
as can be seen in Fig.(\ref{fig:16M_2D_combined}), outputting $Z_{EI}=301$
and $Z_{ET}=86\pm3$. Additional $d=2$ comparisons can be performed
by randomly permuting the count patterns as each permutation allows
one to compare different binned correlation moments. As some correlation
moments may be closer to their theoretical values than others, Z-scores
are likely to vary between permutations, as was observed in threshold
detector experiment comparisons \citep{dellios2025validation}. Focusing
on comparisons with the non-ideal ground truth, a mean increase in
the resulting Z-scores is observed from ten permutations, such that
$\left\langle Z_{ET}\right\rangle _{\mathcal{RP}}\approx139$, where
$\left\langle \dots\right\rangle _{\mathcal{RP}}$ denotes a mean
over random permutations. Although some permutations are closer to
the non-ideal ground truth, with the closest being $Z_{ET}\approx73$,
the majority are further, where the largest observed Z-score was $Z_{ET}\approx235$
from a single permutation.

Although this is not unexpected, such results indicate further systematic
errors besides those tested are present. These may include unequal
losses in each network channel, phase noise, shot-to-shot fluctuations,
miscounted photons, or detector efficiency fluctuations, among others.
We conjecture that the presence of further systematic errors other
than decoherence and measurement errors are also present in the $216$-mode
HS and $288$-mode implementations. However the small sample size
means such errors cannot be resolved in statistical testing for these
data sets, limiting a thorough analysis.

\subsection{Photon number moments}

Finally, for completeness, we simulate the mean output photon number
moments $\left\langle \hat{n}'_{j}\right\rangle $ for all $j$ modes,
where summary tables of statistical test results for all data sets
is also presented in Appendix A. As with the GCP comparisons, count
patterns from the $216$-mode HS and $288$-mode experiments are far
from their theoretical first-order moments for pure squeezed state
inputs with $Z_{EI}\approx86$ and $Z_{EI}\approx192$ for the $216$-mode
HS and $288$-mode data sets respectively.

Unlike the GCP comparisons, decoherence and transmission matrix corrections
no longer cause the theoretical moments to be within the sampling
error-bars of the experimental moments, where $Z_{ET}\approx45$ for
the $216$-mode HS data set and the $288$-mode data set is $Z_{ET}\sigma_{\mathcal{N}}\approx63\sigma_{\mathcal{N}}$
standard deviations away from its expected mean. Similar trends are
present for all data sets where, as was the case for GCPs, the $16$-mode
moments are furthest from their expected first order moments than
any other experiment. We emphasize that statistical testing is more
accurate for this data set due to the large number of experimental
samples, allowing one to resolve finer differences between theoretical
and experimental photon number moments.

\begin{figure}
\begin{centering}
\includegraphics[width=0.5\textwidth]{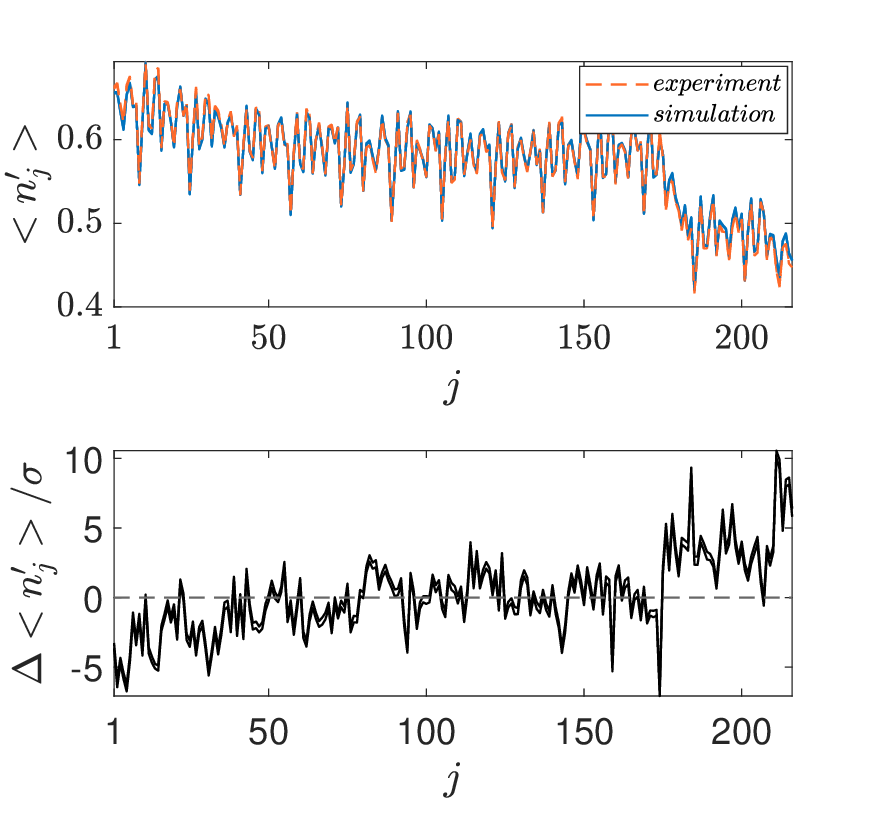}
\par\end{centering}
\caption{\foreignlanguage{australian}{Upper panel: Input thermalised squeezed state theoretical mean output
photon numbers $\left\langle \hat{n}'_{j}\right\rangle $ versus mode
number $j$ are compared to their experimental counterparts for data
from the $216$-mode HS Borealis experiment, where simulations are
performed for an ensemble size of $E_{S}=1.2\times10^{6}$. Decoherence
and $\boldsymbol{T}$-matrix parameters are $\epsilon=0.0510\pm0.0005$
and $t=0.9941\pm0.0005$ which, although improving the fit with experimental
data, no longer cause theory and experiment to agree within sampling
error. Lower panel: Corresponding normalised difference between marginal
moments. \label{fig:216_HS_mean_Pn}}}
\end{figure}

\begin{figure}
\begin{centering}
\includegraphics[width=0.5\textwidth]{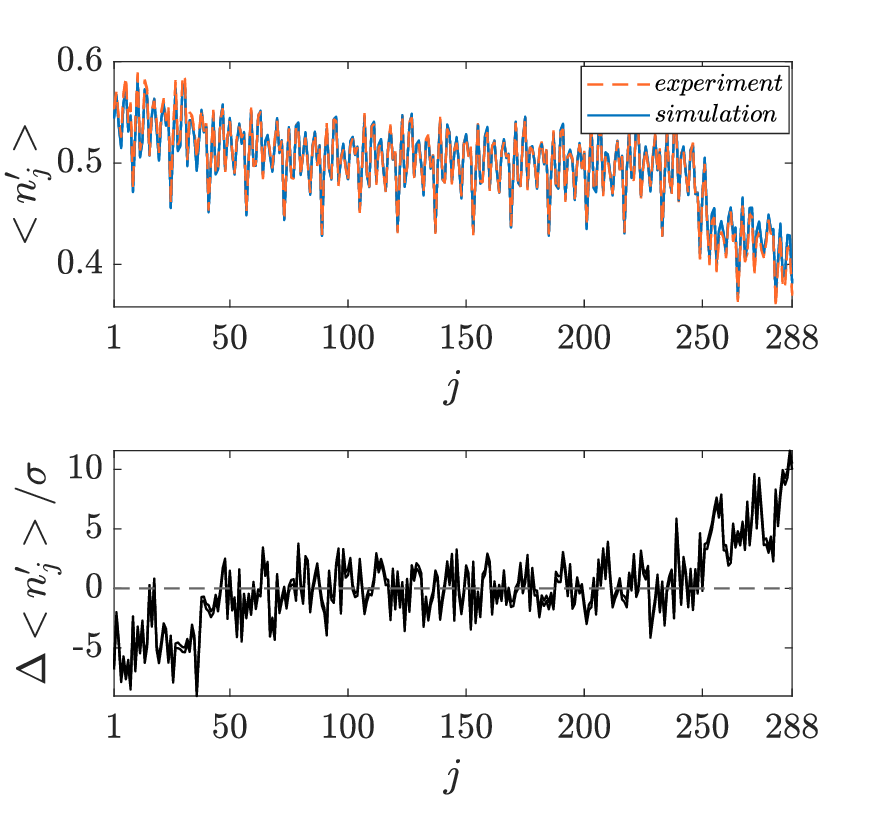}
\par\end{centering}
\caption{\foreignlanguage{australian}{Figure description follows Fig.(\ref{fig:216_HS_mean_Pn}), except
data is now from the $288$-mode Borealis experiment where corresponding
fitting parameters are $\epsilon=0.0547\pm0.0005$ and $t=0.9848\pm0.0005$.
\label{fig:288_HS_mean_Pn}}}
\end{figure}

One noticeable feature in the experimental and theoretical moments
is the decrease in $\left\langle \hat{n}'_{j}\right\rangle $ as $j\rightarrow M$,
with sharp decreases occurring for later modes as seen in Fig.(\ref{fig:216_HS_mean_Pn})
for $175<j<216$ and Fig.(\ref{fig:288_HS_mean_Pn}) for $250<j<288$,
while distinct sharp spikes in the moments occur every $16$ modes.
This clear structure is present in all data sets from this experiment
but is not present in previous implementations with threshold detectors
\citep{zhong2020quantum,zhongPhaseProgrammableGaussianBoson2021}
and is likely due to the time-domain multiplexing scheme used to form
the linear network in Ref. \citep{madsenQuantumComputationalAdvantage2022}.
The experimental set-up interferes a train of input squeezed photons
on multiple fibre delay lines separated by temporal bins of various
dimension. Output pulses are then sent into a 1-to-16 mode demultiplexer
before measurements are performed by PNR detectors.

Considering its input-output ratio the demultiplexer may be the cause
of the periodic spikes. However the structure in the form of the overall
decrease in $\left\langle \hat{n}'_{j}\right\rangle $ for increasing
mode number may be due to the use of fibre delay lines. Although providing
a more compact architecture, it comes at the expense of reduced connectivity,
i.e. photon interference between all modes isn't possible. This has
the effect of introducing a structure to the output photons that can
be exploited by classical samplers such as the tree-width sampler
\citep{ohClassicalSimulationBoson2022}.

\section{Conclusion\label{sec:Conclusion}}

In this paper, we extended a binning method developed by Mandel \citep{mandelCoherencePropertiesOptical1965,Mandel1995_book}
to simulate nonclassical photon statistics of photon-number resolving
detectors using the positive P-representation. We performed validation
tests on quantum computers claiming quantum advantage with photon-number
resolving detectors.

This method is computationally efficient, with computational run-times
of a few minutes on a 12-core desktop at $\sim10^{11}$ floating point
operations/ sec (Linpack Flops) for one-dimensional GCP simulations
of the largest sized networks implemented in the paper. Since the
positive P-representation accurately computes correlation moments
to any order generated in the photonic network, it is a time and resource
efficient method of validation compared to classical algorithms that
compute the Hafnian to replicate photo-detection measurements, or
tensor network approximations \citep{deshpandeQuantumComputationalAdvantage2022a}.

Increasing the GCP dimension causes an increase in computation time,
although this is still significantly faster than other validation
methods. Directly calculating $100\times100$ Hafnians are estimated
\citep{deshpandeQuantumComputationalAdvantage2022a} to take $14$
hours on Fugaku, one of the worlds largest supercomputers in 2024,
at $\sim10^{18}$ Flops. This would compute one probability, but rounding
errors are unknown and are likely to be extremely large at current
$64$ bit numerical precision. Generating enough classical counts
for validation using this approach would take $\sim10^{16}s$ \citep{deshpandeQuantumComputationalAdvantage2022a}
on Fugaku, indicating an overall speed-up of $>10^{18}$ using the
positive-P method at these network sizes.

Our method does not generate classical counts, so it is strictly for
validation of the computational task, and does not replicate the required
random counts. Validation tests were performed on GBS data from a
recent experiment claiming quantum advantage \citep{madsenQuantumComputationalAdvantage2022}.
We showed that within sampling error agreement between theory and
experiment for one-dimensional, but not two-dimensional GCPs is obtained
once decoherence is included in the input states, and a transmission
matrix measurement error correction is applied. There are large discrepancies
from $\chi^{2}$ tests on the individual channel count probabilities,
because these can be calculated and measured very accurately. This
result agrees with a trend obtained previously for threshold detector
experiments \citep{drummondSimulatingComplexNetworks2022,dellios2025validation}.

Comparisons with threshold detector experiments are slightly misleading,
however. Due to the smaller number of experimental samples from data
sets claiming quantum advantage using PNR methods, only a small fraction
of the possible tests could be performed. This was because experimental
sampling errors were too large to apply more sensitive tests. Based
on comparisons of the $16$-mode experimental data, whose sample size
was large enough to perform accurate statistical testing, we conjecture
that more experimental errors are present in the quantum advantage
data sets than those tested for in this paper.

In the longer term, due to its computational efficiency and ability
to generate any measurable moment, the positive-P based binning method
could be used as a means to provide error correction, either by correcting
parameter values as shown in this paper, or by giving feedback so
they can be adjusted. We suggest that this could provide a more practical
route to quantum computational advantage than only using improved
hardware, as is also thought to be the case for logic gate based quantum
computers.

Finally, even though we focused here on GBS networks with pure or
thermalized squeezed states, both the binning method and positive
P-representation can be used to simulate photon statistics of any
nonclassical state.

\section*{Acknowledgments}

ASD would like to thank Nicholas Quesada, Fabian Laudenbach and Jonathan
Lavoie for helpful discussions. This research was funded through grants
from NTT Phi Laboratories and a Templeton Foundation grant ID 62843.

\section*{Appendix A: Summary of all experimental comparisons\label{sec:Summay_exp_comp}}

The Borealis GBS implemented in \citep{madsenQuantumComputationalAdvantage2022}
performed experiments for varying system sizes of $M=16,72,216,\:\text{and}\:288$.
The $216$-mode experiment was repeated for the mean squeezing parameters
$\bar{r}=1.103$ and $\bar{r}=0.533$, hence they are distinguished
as high squeezing (HS) and low squeezing (LS) experiments, respectively.
The ideal ground truth distributions for each data set are formulated
from their corresponding lossy transmission matrices $\boldsymbol{T}$
and simulated pure squeezed states using squeezing vectors $\boldsymbol{r}$.

The graphs presented in Section \ref{sec:GBS_GCP_exp_comp} of the
main text compare theory and experiment predominantly for the $216$-mode
HS and $288$-mode data sets, due to claims of quantum advantage in
the original paper. Here we summarize statistical test results for
all data sets in \citep{madsenQuantumComputationalAdvantage2022},
starting with one-dimensional GCPs for the ideal ground truth in Table.(\ref{tab:GCP_1D_ideal}).
We use the subscripts $EI$ and $ET$ to denote comparisons between
experimental data and the ideal and non-ideal ground truth distributions,
respectively. These results clearly show that there are very large
discrepancies in all data-sets, and that these differences cannot
be explained by the intrinsic sampling error from using a finite number
of samples.

A summary of the non-ideal ground truth comparisons are presented
in Table (\ref{tab:GCP_1D_thermalized}), where the $t$ and $\epsilon$
fitting parameters are found using a Nelder-Mead simplex algorithm
which minimizes the distance between experimental and theoretical
distributions. As stated earlier, experimental data agrees with our
modified theoretical predictions within sampling error for all data
sets bar the $16$-mode experiment.

\begin{table}[H]
\begin{centering}
\begin{tabular}{ccccc}
\toprule 
\multicolumn{2}{c}{Mode number} & $\chi_{EI}^{2}/k$ & $k$ & $Z_{EI}$\tabularnewline
\midrule
\midrule 
\multicolumn{2}{c}{$16$} & $2.53\times10^{3}$ & $37$ & $163$\tabularnewline
\midrule 
\multicolumn{2}{c}{$72$} & $171$ & $56$ & $72$\tabularnewline
\midrule 
\multicolumn{2}{c}{$216$-(LS)} & $165$ & $48$ & $66$\tabularnewline
\midrule 
\multicolumn{2}{c}{$216$-(HS)} & $54$ & $140$ & $70$\tabularnewline
\midrule 
\multicolumn{2}{c}{$288$} & $411$ & $149$ & $167$\tabularnewline
\bottomrule
\end{tabular}
\par\end{centering}
\caption{\foreignlanguage{australian}{Summary of statistical tests for comparisons of one-dimensional GCPs
$\mathcal{G}_{\{1,\dots,M\}}^{(M)}(m)$ for all experimental data
sets obtained from \citep{madsenQuantumComputationalAdvantage2022}.
Chi-square and Z-statistic tests are generated from positive-P simulated
GCPs with an ensemble size of $E_{S}=1.2\times10^{6}$. Although the
$16$-mode and $288$-mode data are furthest from the ideal ground
truth distribution, all experimental count patterns are at least $66\sigma_{\mathcal{N}}$
from their expected normally distributed means. \label{tab:GCP_1D_ideal}}}
\end{table}

\begin{table}[H]
\begin{centering}
\begin{tabular}{ccccccccc}
\toprule 
\multicolumn{4}{c}{Mode number} & $t$ & $\epsilon$ & $\chi_{ET}^{2}/k$ & $k$ & $\left|Z_{ET}\right|$\tabularnewline
\midrule
\midrule 
\multicolumn{4}{c}{$16$} & $0.9841$ & $0.0648$ & $14\pm2$ & $36$ & $18\pm1$\tabularnewline
\midrule 
\multicolumn{4}{c}{$72$} & $0.9861$ & $0.0507$ & $1.02\pm0.03$ & $54$ & $0.2\pm0.2$\tabularnewline
\midrule 
\multicolumn{4}{c}{$216$-(LS)} & $0.9882$ & $0.0105$ & $0.82\pm0.02$ & $47$ & $0.89\pm0.1$\tabularnewline
\midrule 
\multicolumn{4}{c}{$216$-(HS)} & $0.9941$ & $0.0510$ & $0.88\pm0.02$ & $137$ & $1\pm0.2$\tabularnewline
\midrule 
\multicolumn{4}{c}{$288$} & $0.9848$ & $0.0547$ & $0.87\pm0.02$ & $142$ & $1.1\pm0.2$\tabularnewline
\bottomrule
\end{tabular}
\par\end{centering}
\caption{\foreignlanguage{australian}{Description follows those presented in Table.(\ref{tab:GCP_1D_ideal})
except now experimental data is compared to a modified non-ideal ground
truth distribution for the one-dimensional GCPs. Thermalisation component
$\epsilon$ and transmission matrix measurement correction $t$ are
used to modify the ground truth distribution to best fit the experimental
data, where $\epsilon,t$ each have error bars of $\pm0.0005$ for
all data sets. The $\chi_{ET}^{2}/k$ and $Z_{ET}$ values are averages
over ten simulation runs with corresponding standard deviations. \label{tab:GCP_1D_thermalized}}}
\end{table}

Comparisons of the two-dimensional GCPs and mean output photon number
moments are presented in Table.(\ref{tab:GCP_2D}) and Table.(\ref{tab:FO_mom_ideal_thermalized}),
respectively, for the ideal and non-ideal ground truth distributions.
These show large differences in the two-dimensional GCPs compared
to the ideal ground-truth. However, the two-parameter fitting model
used in the one-dimensional comparisons is insufficient to give agreement
in the thermalized case.

\begin{table}[H]
\begin{centering}
\begin{tabular}{cccccccc}
\toprule 
\multicolumn{4}{c}{Mode number} & $Z_{EI}$ & $k$ & $\left|Z_{ET}\right|$ & $k$\tabularnewline
\midrule
\midrule 
\multicolumn{4}{c}{$16$} & $301$ & $426$ & $86\pm3$ & \selectlanguage{australian}%
$416$\selectlanguage{american}%
\tabularnewline
\midrule 
\multicolumn{4}{c}{$72$} & $82$ & $772$ & \selectlanguage{australian}%
$12\pm1$\selectlanguage{american}%
 & \selectlanguage{australian}%
$736$\selectlanguage{american}%
\tabularnewline
\midrule 
\multicolumn{4}{c}{$216$-(LS)} & $75$ & $658$ & \selectlanguage{australian}%
$10\pm1$\selectlanguage{american}%
 & \selectlanguage{australian}%
$640$\selectlanguage{american}%
\tabularnewline
\midrule 
\multicolumn{4}{c}{$216$-(HS)} & $58$ & $4326$ & \selectlanguage{australian}%
$9\pm1$\selectlanguage{american}%
 & \selectlanguage{australian}%
$4154$\selectlanguage{american}%
\tabularnewline
\midrule 
\multicolumn{4}{c}{$288$} & $203$ & $4877$ & \selectlanguage{australian}%
$11\pm1$\selectlanguage{american}%
 & \selectlanguage{australian}%
$4571$\selectlanguage{american}%
\tabularnewline
\bottomrule
\end{tabular}
\par\end{centering}
\caption{\foreignlanguage{australian}{Summary of statistical tests for comparisons of two-dimensional GCPs
$\mathcal{G}_{\{1,\dots,M/2\},\{M/2+1,\dots,M\}}^{(M)}(m_{1},m_{2})$
for all experimental data sets obtained from \citep{madsenQuantumComputationalAdvantage2022}.
Chi-square and Z-statistic tests are generated from positive-P simulated
GCPs with an ensemble size of $E_{S}=1.2\times10^{6}$. The $\epsilon$
and $t$ fitting parameters are the same those used in Table.(\ref{tab:GCP_1D_thermalized}),
where $Z_{ET}$ values are averages over ten simulation runs with
corresponding standard deviations. \label{tab:GCP_2D}}}
\end{table}

\begin{table}[H]
\begin{centering}
\begin{tabular}{cccc}
\toprule 
\multicolumn{2}{c}{Mode number} & $Z_{EI}$ & $Z_{ET}$\tabularnewline
\midrule
\midrule 
\multicolumn{2}{c}{$16$} & $264$ & $243$\tabularnewline
\midrule 
\multicolumn{2}{c}{$72$} & $78$ & $17$\tabularnewline
\midrule 
\multicolumn{2}{c}{$216$-(LS)} & $85$ & $37$\tabularnewline
\midrule 
\multicolumn{2}{c}{$216$-(HS)} & $86$ & $45$\tabularnewline
\midrule 
\multicolumn{2}{c}{$288$} & $192$ & $63$\tabularnewline
\bottomrule
\end{tabular}
\par\end{centering}
\caption{\foreignlanguage{australian}{Summary of Z-statistic test results for comparisons of both the ideal
and non-ideal ground truth distributions for the mean output photon
number $\left\langle \hat{n}'_{j}\right\rangle $ of each experimental
data set. Positive-P simulations are performed for an ensemble size
of $E_{S}=1.2\times10^{6}$ for all data sets except the $16$-mode
experiment, where $E_{S}=3.36\times10^{7}$ is needed in order for
$\sigma_{T,j}\apprle\sigma_{E,j}$ which is required for accurate
simulations. Fitting parameters applied to each data set are obtained
from Table.(\ref{tab:GCP_1D_thermalized}). \label{tab:FO_mom_ideal_thermalized}}}
\end{table}

The last table, giving the comparisons for the mean photon number
per channel, has a much lower sampling error than the other results
tabulated, since all the output samples can be used to obtain these
mean values. This shows discrepancies outside the sampling error limits,
indicating that transmission matrix parameter errors could be a significant
part of the overall differences. This is consistent with the increased
errors found for the two-dimensional GCPs.

\section*{Appendix B: Estimating sampling errors of moments}

Accurately estimating sampling errors is essential when comparing
theoretical simulations with experimental data obtained via sampling.
This is particularly important when using statistical tests that are
dependent on sampling errors, as incorrect estimates will result in
misleading test results.

Theoretical sampling errors are estimated numerically by subdividing
the total phase-space ensemble as $E_{S}=N_{S}N_{R}$. Breaking the
full ensemble of phase-space samples into sub-ensembles is a common
numerical procedure for applications of phase-space representations
yielding two complementary benefits: efficient multi-core parallel
computing, decreasing computation time depending on available resources,
and estimates of theoretical sampling errors \citep{opanchuk2018simulating}.

The first sub-ensemble corresponds to sampling from the Gaussian phase-space
distribution $N_{S}$ times and $N_{R}$ is the number of times the
computation is repeated. Provided $N_{R}\gg1$, the second sub-ensemble
allows one to accurately estimate the standard deviation in the ensemble
mean as $\sigma_{T,i}=\sigma_{t,i}/\sqrt{N_{R}}$ \citep{freundStatisticalMethods2003,kloedenStochasticDifferentialEquations1992},
where $\sigma_{t}$ is the computed sample standard deviation which
takes the general form \citep{opanchuk2018simulating}
\begin{equation}
\sigma_{t}=\sqrt{\frac{1}{N_{R}-1}\sum_{i=1}^{N_{R}}\left(\bar{O}^{(i)}-\bar{O}\right)}.
\end{equation}
Here, $O(\boldsymbol{\alpha},\boldsymbol{\beta})$ is some general
phase-space observable, $\bar{O}^{(i)}=N_{S}^{-1}\sum_{k=1}^{N_{S}}(O^{(k)})^{(i)}$
denotes the first sub-ensemble mean for the $i\in N_{R}$ trajectory
and $\bar{O}=E_{S}^{-1}\sum_{e=1}^{E_{S}}O_{(e)}$ is the full ensemble
mean. In statistics literature \citep{freundStatisticalMethods2003,raoLinearStatisticalInference2009},
$\sigma_{T,i}$ is commonly called the standard error of the mean
and its valid for both binned distributions and moments.

For experimental sampling errors defined as $\sigma_{E,i}=\sigma_{e,i}/\sqrt{N_{E}}$,
we are interested in estimating the standard deviation of the experimental
sample means $\sigma_{e,i}$. Unlike theoretical sampling errors this
isn't performed numerically. Instead, the variances of GCPs for both
threshold and PNR detector experiments have Poissonian fluctuations,
such that $\sigma_{e,i}=\sqrt{\mathcal{G}_{S,i}}$ for some general
input state $S$. Since the true theoretical GCP $\mathcal{G}_{S,i}$
is seldom known, we estimate it in phase-space as $\mathcal{G}_{S,i}=\lim_{E_{S}\rightarrow\infty}\bar{\mathcal{G}}_{S,i}$.

However, estimating $\sigma_{e,i}$ for moments (either raw or central)
and cumulants is somewhat more complicated as these aren't probabilities
but instead tell one about various properties of a probability distribution.
In this appendix, we describe the method of estimating experimental
sampling errors of raw moments used in the comparisons presented in
the main text. Details on how to estimate sampling errors for central
moments and cumulants can be found in \citep{raoLinearStatisticalInference2009,fisherMomentsProductMoments1930,mccullaghTensorMethodsStatistics2018}.

\subsection*{Sampling errors of raw moments}

Let $Y$ be an i.i.d random variable with probability density function
$P(y)$, then the $n$-th raw moment is defined as
\begin{equation}
\mu'_{n}=\left\langle Y^{n}\right\rangle =\int y^{n}P(y)\text{d}y.\label{eq:raw_moment}
\end{equation}

Suppose one generates $\mathcal{S}$ samples denoted $y_{1},\dots,y_{\mathcal{S}}$
from the distribution $P(y)$. The $n$-th sample raw moment is then
defined as \citep{raoLinearStatisticalInference2009,fisherMomentsProductMoments1930,mccullaghTensorMethodsStatistics2018}
\begin{equation}
m'_{n}=\frac{1}{\mathcal{S}}\sum_{i=1}^{\mathcal{S}}y_{i}^{n},\label{eq:SRM}
\end{equation}
the expected value of which is an unbiased estimator of $\mu'_{n}$
\begin{equation}
\left\langle m'_{n}\right\rangle =\frac{1}{\mathcal{S}}\sum_{i=1}^{\mathcal{S}}\left\langle y_{i}^{n}\right\rangle =\mu'_{n}.\label{eq:SRM_expected_value}
\end{equation}

For our purposes, we are interested in the variance $\sigma_{m'_{n}}^{2}=\left\langle (m'_{n})^{2}\right\rangle -\left\langle m'_{n}\right\rangle ^{2}$.
Following from Ref. \citep{raoLinearStatisticalInference2009}, the
expectation value of the squared moment is simplified as 
\begin{align}
\left\langle (m'_{n})^{2}\right\rangle  & =\frac{1}{\mathcal{S}^{2}}\left\langle \sum_{i=1}^{\mathcal{S}}\sum_{j=1}^{\mathcal{S}}y_{i}^{n}y_{j}^{n}\right\rangle \nonumber \\
 & =\frac{1}{\mathcal{S}^{2}}\left\langle \sum_{l=1}^{\mathcal{S}}y_{l}^{2n}+\sum_{i\neq j}^{\mathcal{S}}y_{i}^{n}y_{j}^{n}\right\rangle ,
\end{align}
where the summations have been expanded to include instances of equal
$l=i=j$ and non-equal $i\neq j$ samples. Utilizing the simplifications
$\sum_{i=1}^{\mathcal{S}}\left\langle y_{i}^{n}\right\rangle =\mathcal{S}\mu'_{n}$
from Eq.(\ref{eq:SRM_expected_value}) and $\sum_{i\neq j}^{\mathcal{S}}=\sum_{i}^{\mathcal{S}-1}\sum_{j}^{\mathcal{S}}$,
the second-order sample raw moment is simplified as
\begin{equation}
\left\langle (m'_{n})^{2}\right\rangle =\frac{\mu'_{2n}+(\mathcal{S}-1)(\mu'_{n})^{2}}{\mathcal{S}},
\end{equation}
allowing one to estimate the variance in the sample raw moment as
\citep{raoLinearStatisticalInference2009}
\begin{equation}
\sigma_{m'_{n}}^{2}=\frac{\mu'_{2n}-(\mu'_{n})^{2}}{\mathcal{S}}.\label{eq:SRM_variance}
\end{equation}

Its possible to extend the above description for univariate statistics
to the multivariate case. Let $\boldsymbol{Y}=[Y_{1},Y_{2},\dots]$
be a vector of i.i.d random variables with multivariate probability
density $P(\boldsymbol{y})$, where $\boldsymbol{y}=[y_{1},y_{2},\dots]$.
The $n$-th raw moment is then defined as 
\begin{equation}
\mu'_{n_{1},n_{2},\dots}=\left\langle Y_{1}^{n_{1}}Y_{2}^{n_{2}}\dots\right\rangle =\int\left(y_{1}^{n_{1}}y_{2}^{n_{2}}\dots\right)P(\boldsymbol{y})\text{d}\boldsymbol{y}.
\end{equation}
As in the univariate case, suppose one generates $\mathcal{S}$ samples
of the $j=1,2,\dots$-th random variable in $\boldsymbol{Y}$, denoted
$y_{1,j},\dots,y_{\mathcal{S},j}$, then the multivariate $n$-th
sample raw moment is defined as \citep{fisherMomentsProductMoments1930}
\begin{equation}
m'_{n_{1},n_{2},\dots}=\frac{1}{\mathcal{S}}\sum_{i=1}^{\mathcal{S}}y_{i,1}^{n_{1}}y_{i,2}^{n_{2}}\cdots,
\end{equation}
whose expected value is an unbiased estimate of the multivariate raw
moment \citep{mccullaghTensorMethodsStatistics2018}
\begin{equation}
\left\langle m'_{n_{1},n_{2},\dots}\right\rangle =\frac{1}{\mathcal{S}}\sum_{i=1}^{\mathcal{S}}\left\langle y_{i,1}^{n_{1}}y_{i,2}^{n_{2}}\cdots\right\rangle =\mu'_{n_{1},n_{2},\dots}.
\end{equation}

Following from the derivation of the univariate case, the variance
of the multivariate raw sample moment is obtained as \citep{kendaladstat}
\begin{equation}
\sigma_{m'_{n_{1},n_{2},\dots}}^{2}=\frac{\mu'_{2n_{1},2n_{2},\dots}-(\mu'_{n_{1},n_{2},\dots})^{2}}{\mathcal{S}}.
\end{equation}
By setting $n_{1}=n_{2}=\dots=1$, sampling errors for higher photon
number moments such as $\left\langle \hat{n}'_{j}\hat{n}'_{k}\right\rangle $,
$\left\langle \hat{n}'_{j}\hat{n}'_{k}\hat{n}'_{l}\right\rangle $
can now be estimated.

\subsection*{Accuracy of sampling error estimates}

In this paper, the raw moments of interest are the mean output photon
numbers per mode $\left\langle \hat{n}'_{j}\right\rangle $. Using
Eq.(\ref{eq:SRM_variance}), the experimental sampling errors can
therefore be estimated as 
\begin{equation}
\sigma_{E,j}=\frac{\sigma_{e,j}}{\sqrt{N_{E}}}=\sqrt{\frac{\left\langle \hat{n}_{j}^{\prime2}\right\rangle -\left\langle \hat{n}'_{j}\right\rangle ^{2}}{N_{E}}},\label{eq:Exp_samp_error_moments}
\end{equation}
where, as was the case for GCPs, we estimate the true theoretical
photon number moments in phase-space as $\left\langle \hat{n}'_{j}\right\rangle =\lim_{E_{S}\rightarrow\infty}\bar{n}'_{j}$.
To determine the accuracy of this estimate, we must generate photon
count patterns corresponding to PNR detectors from an exactly known
model and compare the estimated sampling error with the standard deviation
of the mean output photon number of the count patterns from their
exact values. To do this, the simplest method is to use classical
states.

\subsubsection*{Classical state PNR photo-detection algorithm}

In the case of $N$ classical states input into a GBS linear network,
one has rotated to a classical phase-space and we can use the coherent
amplitudes corresponding to the diagonal P-representation to replicate
photo-detection measurements of PNR detectors.

Generally, each stochastic trajectory in $E_{S}$ corresponds to a
single experimental shot. Therefore one can conserve the classical
correlations generated within the network by computing the phase-space
observable projector of the $j$-th detector $p_{j}$, which corresponds
to the operator Eq.(\ref{eq:PNR_detector_projector}), for the $k$-th
trajectory as
\begin{equation}
(p_{j}(c_{j}^{f}))^{(k)}=\frac{1}{c_{j}^{f}!}\left(\left(n'_{j}\right)^{c_{j}^{f}}e^{-n'_{j}}\right)^{(k)},
\end{equation}
where $k\in E_{S}$, $n'_{j}=\left|\alpha_{j}\right|^{2}$ in the
diagonal P-representation and $c_{j}^{f}=0,1,2,\dots,c_{j}^{(\text{max}),f}$
is a photon count of our fake experiment up to some maximum cut-off
$c_{j}^{(\text{max}),f}$.

The phase-space projectors determine the probability of observing
a specific count $c_{j}^{f}$. Since each detector can distinguish
between multiple oncoming photons, the actual computation involves
calculating a vector of probabilities for each stochastic trajectory:
\begin{equation}
\boldsymbol{P}_{j}^{(k)}=[P_{0,j}^{(k)},P_{1,j}^{(k)},\dots,P_{\text{max},j}^{(k)}],\label{eq:PNR_fake_probability_vector}
\end{equation}
where $P_{0,j}^{(k)}=(p_{j}(0))^{(k)}$ is the success probability
of observing no counts at the $j$-th detector, $P_{1,j}^{(k)}=(p_{j}(1))^{(k)}$
is the success probability of observing a single count at the $j$-th
detector, and so on. The PNR photo-detection measurements are then
replicated by randomly sampling Eq.(\ref{eq:PNR_fake_probability_vector}),
which outputs the fake photon count vector 
\[
(\boldsymbol{c}^{f})^{(k)}=[P_{c_{1},1}^{(k)},P_{c_{2},2}^{(k)},\dots,P_{c_{M},M}^{(k)}].
\]

As the method scales with the total phase-space ensemble size, the
number of faked patterns generated is $N_{F}=E_{S}$. These are binned
in the same way as the experimental data, where $m_{j}^{f}=\sum_{i\in S_{j}}c_{i}^{f}$
is the fake grouped count corresponding to the GCP $\mathcal{G}_{j}^{f}=N_{F}^{-1}\sum_{i\in S_{j}}c_{i}^{f}$.

\subsubsection*{Tests of estimate accuracy}

The photon counting distributions for a number of classical states
can be exactly computed, one of which is multi-mode thermal states.
The one-dimensional GCP of a multi-mode thermal state is obtained
from the single-mode distribution \citep{walls2008quantum}
\begin{equation}
\mathcal{G}_{S}^{(M)}(m)=p(1-p)^{m},
\end{equation}
where $p=1/(1+n)$ is the success probability requiring a uniform
squeezing parameter and hence equal photon numbers $n=n_{1}=\dots=n_{N}$.
The single-mode theory follows a geometric distribution which is a
special case of the negative binomial distribution: if $X$ is an
independent and geometrically distributed random variable with parameter
$p$, then $Y=\sum_{j}^{M}X_{j}$ follows a negative binomial distribution
with parameters $M$ and $p$.

The one-dimensional GCP theory of multi-mode thermal state inputs
therefore follows a negative binomial distribution 
\begin{equation}
\mathcal{G}_{S}^{(M)}(m)=\left(\begin{array}{c}
m+M-1\\
m
\end{array}\right)p^{M}(1-p)^{m}.
\end{equation}
For the case of thermal photons transformed by a Haar random unitary,
the input and output mean photon numbers are equal $n=\left\langle \hat{n}'\right\rangle =\sinh^{2}(r)$
and the photon number variance can be computed exactly as $\sigma_{n}^{2}=n\left(n+1\right)=\sinh^{2}(r)\left(\sinh^{2}(r)+1\right)$
\citep{loudonQuantumTheoryLight1983}.

\begin{figure}[H]
\begin{centering}
\includegraphics[width=0.5\textwidth]{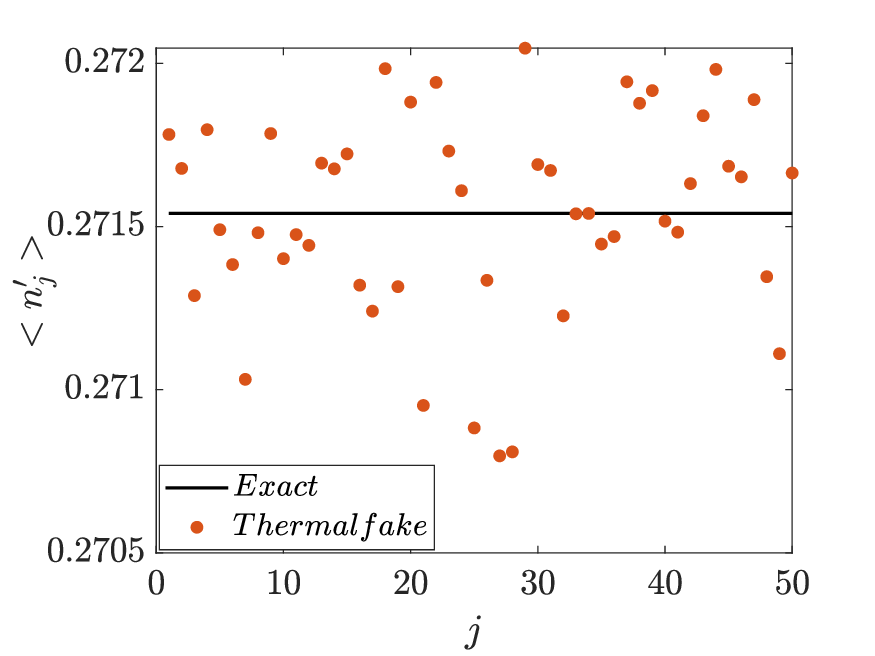}
\par\end{centering}
\caption{\foreignlanguage{australian}{The exact mean photon number (solid black line) of $N=50$ thermal
states sent into a lossless linear network compared to the mean photon
number (solid orange dots) of a classical algorithm that replicates
PNR measurements of classical states using the diagonal P-representation.
The $N_{F}=4\times10^{6}$ count patterns are binned to form the observable
moments, and the standard deviation of these moments from their exact
values $\sigma_{A}$ is the sampling error estimated using Eq.(\ref{eq:thermal_sampling_error_estimate}).\label{fig:Estimate_test_thermal_fake_exact}}}
\end{figure}

Now that we have a method of generating photon counts and an exactly
known model with which to compare those counts, we calculate the accuracy
of the estimate Eq.(\ref{eq:Exp_samp_error_moments}) for $M=N=50$
thermal states with uniform squeezing $\boldsymbol{r}=[0.5,\dots,0.5]$
sent into a lossless linear network described by a Haar random unitary
matrix $\boldsymbol{U}$. We generate $N_{F}=4\times10^{6}$ count
patterns using a detector cut-off of $c_{j}^{(\text{max}),f}=13$
and bin them to compute the mean output photon number $\bar{n}_{j}^{\prime f}$,
which is compared to the exact value in Fig.(\ref{fig:Estimate_test_thermal_fake_exact}).

Following from Eq.(\ref{eq:Exp_samp_error_moments}), the sampling
error of our fake experiment is estimated by setting $\sigma_{e}=\sigma_{n}=\sqrt{n(n+1)}$,
such that
\begin{equation}
\sigma_{E}=\sqrt{\frac{n(n+1)}{N_{F}}}.\label{eq:thermal_sampling_error_estimate}
\end{equation}
If this is an accurate estimate, it should converge to the actual
error between moments
\begin{equation}
\sigma_{A}=\sqrt{\frac{1}{M}\sum_{j=1}^{M}\left(\bar{n}_{j}^{\prime f}-n\right)^{2}},
\end{equation}
which is simply the standard deviation as $n$ is our expected mean
value.

For the data in Fig.(\ref{fig:Estimate_test_thermal_fake_exact}),
our estimated sampling error is $\sigma_{E}\approx2.938\times10^{-4}$
while the actual error is $\sigma_{A}\approx3.095\times10^{-4}$,
corresponding to an $\approx95\%$ agreement. If instead we assumed
one could estimate the errors as Poissonian fluctuations, such as
for GCPs, then $\sigma_{E}=\sqrt{n/N_{F}}\approx2.605\times10^{-4}$
which is only $\approx84\%$ of the actual error.

Therefore, Eq.(\ref{eq:Exp_samp_error_moments}) is an accurate method
of estimating experimental sampling errors of moments. Naturally,
the larger the number of experimental samples the better the estimate
will be as the experimental moments will converge to their ground
truth values more readily.

\bibliography{bosonsamplingPNR}

\begin{thebibliography}{60}
\bibitem [{Pri()}]{PrivateCommunicationPhoton}%
  \BibitemOpen
  \href@noop {} {\bibinfo {title} {Private communication. {{Photon}} counting distribution for the ideal ground truth of the 288-mode data was independently cross-validated. {{The}} distribution presented in the {{Borealis}} paper is closer to the experimental data than the actual ideal ground truth presented here.}}\BibitemShut {Stop}%
\end{thebibliography}
\selectlanguage{australian}%

\end{document}